\documentclass[12pt]{iopart}
\usepackage{graphicx}
\usepackage{bm}
\usepackage{amstext,amsgen,amssymb,amsfonts,amsbsy}
\usepackage{aas_macros}
\usepackage{url}
\usepackage{wasysym}
\usepackage[capitalise]{cleveref}
\usepackage[dvipsnames]{xcolor}
\usepackage{mathrsfs}
\newcommand{\w}[1]{\boldsymbol{#1}}

\newcommand{\hold}{h^{\text{old}}}
\newcommand{\hnew}{h^{\text{new}}}

\newcommand{\beq}{\begin{equation}}
\newcommand{\eeq}{\end{equation}}
\newcommand{\beqn}{\begin{eqnarray}}
\newcommand{\eeqn}{\end{eqnarray}}

\begin{document}
\vspace{-2.5cm}

\title{Improvements to the construction of binary black hole initial data}

\author{Serguei Ossokine$^{1,2}$, Francois Foucart$^{1,3}$, Harald
  P. Pfeiffer$^{1,4}$, Michael Boyle$^{5}$, B\'{e}la Szil\'{a}gyi$^{6}$}

\address{${}^{1}$Canadian Institute for Theoretical Astrophysics,
  University~of~Toronto, Toronto, Ontario M5S 3H8, Canada}
\address{${}^{2}$Department of Astronomy and Astrophysics, 50 St.\
  George Street, University of Toronto, Toronto, ON M5S 3H4, Canada}
\address{${}^{3}$Lawrence Berkeley National Laboratory, 1 Cyclotron
  Rd, Berkeley, CA 94720, USA; Einstein Fellow }
\address{${}^{4}$Canadian Institute for Advanced Research, 180 Dundas
  St.~West, Toronto, ON M5G 1Z8, Canada} \address{${}^{5}$ Center for
  Radiophysics and Space Research, Cornell University, Ithaca, New
  York, 14853} \address{${}^{6}$Theoretical Astrophysics 350-17,
  California Institute of Technology, Pasadena, CA 91125}

\date{\today}

\begin{abstract}

  Construction of binary black hole initial data is a prerequisite for
  numerical evolutions of binary black holes.  This paper reports
  improvements to the binary black hole initial data solver in the
  Spectral Einstein Code, to allow robust construction of initial data
  for mass-ratio above 10:1, and for dimensionless black hole spins
  above 0.9, while improving efficiency for lower mass-ratios and
  spins.  We implement a more flexible domain decomposition, adaptive
  mesh refinement and an updated method for choosing free
  parameters. We also introduce a new method to control and eliminate
  residual linear momentum in initial data for precessing systems,and
  demonstrate that it eliminates gravitational mode mixing during the
  evolution. Finally, the new code is applied to construct initial data
  for hyperbolic scattering and for binaries with very small
  separation.
\end{abstract}


\submitto{\CQG}

%%%%%%%%%%%%%%%%%%%%%%%%%%%%%%%%%%%%%%%%%%%%%%%%%%%%%%%%%%%%%%%%
%%%%%%%%%%%%%%%%%%%%%%%%%%%%%%%%%%%%%%%%%%%%%%%%%%%%%%%%%%%%%%%%
\section{Introduction}
\label{s:Introduction}
%%%%%%%%%%%%%%%%%%%%%%%%%%%%%%%%%%%%%%%%%%%%%%%%%%%%%%%%%%%%%%%%
%%%%%%%%%%%%%%%%%%%%%%%%%%%%%%%%%%%%%%%%%%%%%%%%%%%%%%%%%%%%%%%%

Almost a century ago the existence of gravitational waves was first
predicted~\cite{Einstein1918}. Gravitational radiation offers an
exciting new observational
window~\cite{Sathyaprakash:2009xs,Hobbs:2009yy} and the enticing
possibility of multimessenger astronomy. With the second generation of
gravitational wave detectors poised to come online~\cite{aLIGO1,
  ALIGO2, AdV}, it is more important than ever to model the likely
sources of gravitational waves. Some of the most promising are binary
black holes, with predicted detection rates of $0.4-1000$ per year for
Advanced LIGO~\cite{Abadie:2010cfa}.  To detect such systems, matched
filtering techniques must be used in order to extract the signal from
the noise~\cite{Finn:1992}. This requires accurate models of binary
black hole inspiral, merger and ringdown.  A vast amount of work has
been done in this direction in full numerical relativity which is
necessary to describe the very dynamic plunge and merger regimes (see
e.g.~\cite{2014ARA&A..52..661L,Pfeiffer:2012pc,Hinder:2010vn,McWilliams:2010iq,Hannam:2009rd}
for overviews of the field).  While many groups now successfully
simulate binary black hole
systems~\cite{Pekowsky:2013ska,ninja,Ajith:2012az,Ajith:2012az-PublicData,Hinder:2013oqa},
much of the vast 7-dimensional parameter space consisting of the mass
ratio $q$ and the dimensionless spins $\w{\chi}_{A,B}$ remains
unexplored. Most of the attention has been focused on binaries close
to equal mass ($q\lesssim8$) and modest spin
($\chi_{A,B}\lesssim 0.8$) (although
see~\cite{LoustoZlochower2010,Sperhake:2011ik,Scheel2014,Varma:2014,Bustillo:2015ova,Shoemaker:2015cea})
For stellar mass black hole binaries, one can expect mass ratios
$\lesssim 15$ and arbitrary spin magnitudes and orientations, which
leads to precession of the spins and the orbital plane.  Precessing,
high mass-ratio binaries have interesting dynamics, causing large
modulations of the gravitational waveform. One can expect even higher
mass ratios ($q\simeq 30$) for neutron star-black hole (NSBH) binaries
(see~\cite{2008ApJ...678L..17S} a BH-Wolf-Rayet system with BH mass
$~30M_{\astrosun}$). At high mass ratios, BBH systems can be used as
proxies for NSBH systems(e.g. ~\cite{Foucart:2013psa}). One would thus
like to simulate high-mass ratio BBH systems.

Intermediate mass black holes (IMBH) with masses
$m=10^{2}-10^{4} M_{\astrosun}$ have been hypothesised to exist to
complete the BBH mass hierarchy (e.g., the
review~\cite{2004IJMPD..13....1C}). Searches for IMBH have been
performed and several candidates have been identified (see
e.g.~\cite{Mezcua01042015,Pasham2014} for recent observations). Higher
mass ratio ($10\lesssim q\lesssim 100$) systems may serve as models
for binaries containing an IMBH and a stellar mass black hole or
neutron star.  Advanced era gravitational wave detectors might be able
to observe gravitational waves from such systems, with a detection
rate of up to 10 events per year for stellar-mass - IMBH
binaries~\cite{Abadie:2010cfa}.  It is thus important to explore these
systems in numerical relativity.

The first step to numerically evolving a binary black hole spacetime
is the construction of appropriate data on the initial
hypersurface~\cite{Cook2000}.  This involves the solution of the
elliptic constraint equations with free data that corresponds to a
binary in quasi-equilibrium, ideally allowing for arbitrary masses,
spins and velocities of the two black holes.  The Spectral Einstein
Code ({\tt SpEC})~\cite{SpECwebsite} includes a BBH initial data
solver~\cite{Pfeiffer2003} based on the extended conformal thin
sandwich equations~\cite{York1999,Pfeiffer2003b}, incorporating
quasi-equilibrium black hole boundary
conditions~\cite{Cook2002,Cook2004,Caudill-etal:2006}.  This solver
has been used to construct BBH for a wide range of
configurations~\cite{Mroue:2013PRL}.  Construction of BBH with
increasing mass-ratio, increasing spin magnitudes and the desire to
construct initial data for highly spinning BBH with arbitrary spin
axes have necessitated a variety of improvements to the initial data
code compared to its original
presentation~\cite{Pfeiffer2003,Cook2004,Pfeiffer-Brown-etal:2007,Lovelace2008,Buchman:2012dw}.

This paper summarizes these improvements and extends the original code
even further, in anticipation of future desire to study even more
generic BBH systems.  Specifically, here, we
present:\\
(i) Flexible domain-decomposition to allow a wider range of
mass-ratios, spins and separations.\\
(ii) Adaptive mesh-refinement to enhance computational efficiency and
to ensure robust numerical convergence
for mass-ratios $q\gtrsim 5$ and dimensionless spins $\gtrsim 0.9$. \\
(iii) Improved updating formulae for iterative determination of the
free parameters. These formulas allow one to achieve very high spins
and mass ratios, for example an equal-mass binary with aligned spins
of $0.9999$, and a $q=50$ single-spinning binary with
spin of $0.95$ on the large black hole.  \\
(iv) Building on previous work
~\cite{Foucart:2010eq,Henriksson:2014tba}, we control of the ADM
linear momentum to avoid drifts of the center of mass in BBH
evolutions. This eliminates gravitational mode mixing due to the
motion of the centre
of mass with respect to a fixed extraction sphere.\\
(v) Control of the center of mass.

This paper is organized as follows.  In
Sec.~\ref{sec:NumericalTechniques} we describe in detail the numerical
enhancements and additions to the code. In
Sec.~\ref{sec:NumericalResults} we present the results of initial data
construction for several challenging configurations as well as an
exploratory evolution of a new data set that demonstrates that the
control of linear momentum in initial data leads to the elimination of
gravitational wave mode mixing. Finally we summarize the results in
Sec.~\ref{sec:Discussion} and introduce the construction of initial
data for closely separated binaries and binaries on hyperbolic orbits
as applications of the techniques developed in this paper.  

%%%%%%%%%%%%%%%%%%%%%%%%%%%%%%%%%%%%%%%%%%%%%%%%%%%%%%%%%%%%%%%%
%%%%%%%%%%%%%%%%%%%%%%%%%%%%%%%%%%%%%%%%%%%%%%%%%%%%%%%%%%%%%%%%
\section{Numerical techniques}
\label{sec:NumericalTechniques}
%%%%%%%%%%%%%%%%%%%%%%%%%%%%%%%%%%%%%%%%%%%%%%%%%%%%%%%%%%%%%%%%
%%%%%%%%%%%%%%%%%%%%%%%%%%%%%%%%%%%%%%%%%%%%%%%%%%%%%%%%%%%%%%%%

The main task of constructing initial data is twofold: first, to solve
the elliptic constraint equations on the initial hypersurface; and
then, to ensure that the solution represents the astrophysical
situation of interest (in our case, a black-hole binary in
quasi-equilibrium). In {\tt SpEC}, the former is achieved by using a
pseudo-spectral multidomain method; see~\cite{Pfeiffer2003}. The
number of subdomains is kept fixed, but the resolution of each
subdomain is dynamically adjusted to obtain low truncation error.  To
enforce quasi-equilibrium conditions, {\tt SpEC} employs the extended
conformal thin sandwich (XCTS) formalism~\cite{Pfeiffer2003b}. Before
solving the conformal thin-sandwich equations, various free parameters
must be chosen - for example, the sizes of the excision regions, 
and certain other parameters that
affect the location, spin or motion of the black holes.  The free parameters
differ from the physical parameters one desires to control, such as
the masses and spins of the black holes, or the linear momentum
$\mathbf{P}_{\rm ADM}$ of the initial data hypersurface.  Therefore,
iterative root-finding is needed, as described in Buchman et
al~\cite{Buchman:2012dw}.  To minimize the computational cost
associated with many iterations of high resolution solves, we adopt a hybrid approach. The
resolution of the domain and the free parameters are adjusted
simulateneously based on the current estimated truncation error and
the differences between the desired and obtained physical quantities.

In the remainder of this section, we describe in detail the
improvements to the initial data code.

%%%%%%%%%%%%%%%%%%%%%%%%%%%%%%%%%%%%%%%%%%%%%%%%%%%%%%%%%%%%%%%%
\subsection{Domain decomposition}
\label{sec:Domain}
%%%%%%%%%%%%%%%%%%%%%%%%%%%%%%%%%%%%%%%%%%%%%%%%%%%%%%%%%%%%%%%%

\begin{figure}
\centerline{\includegraphics[width=0.8\textwidth]{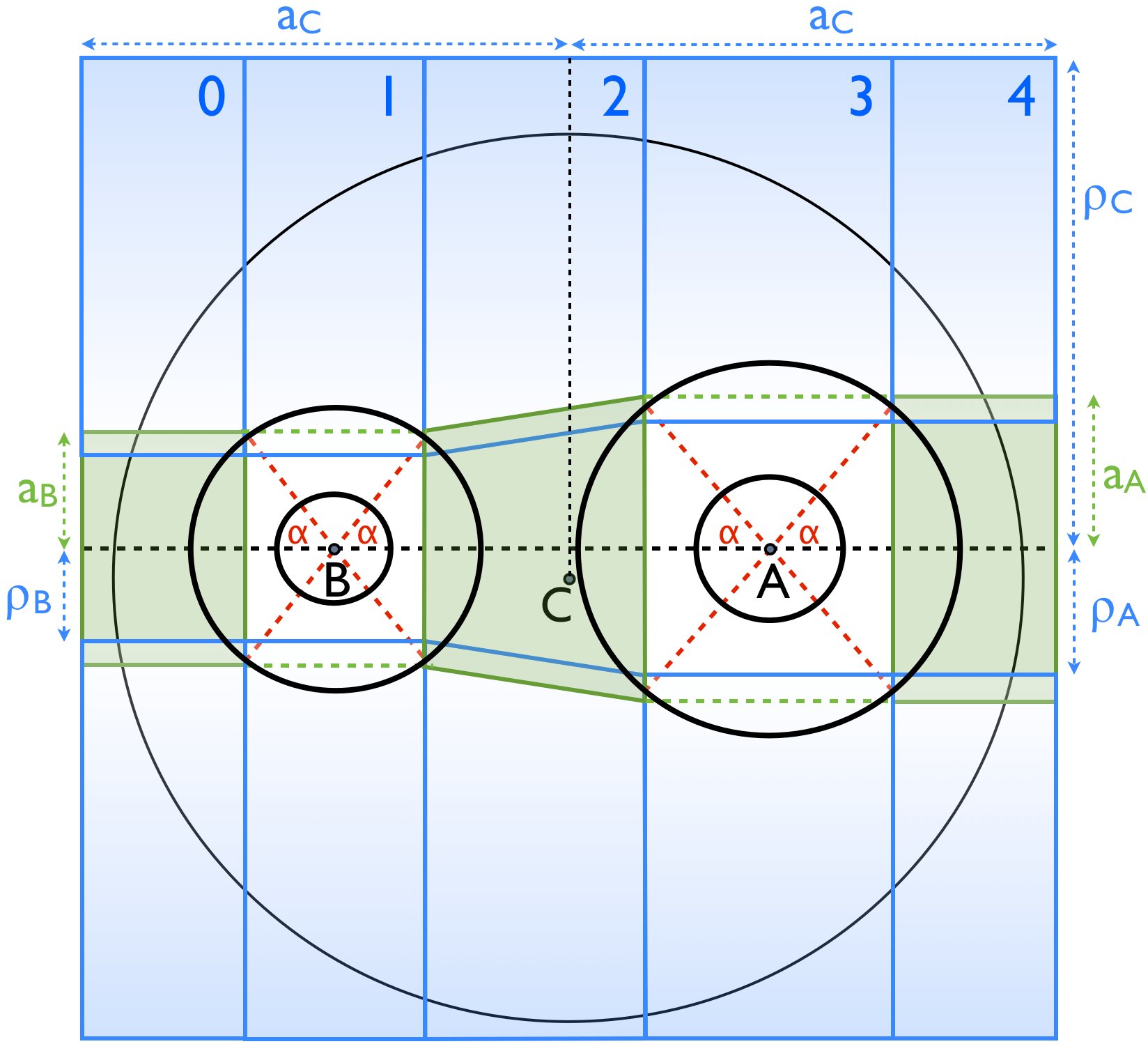}}
\caption{\label{fig:Domain} Schematic of the domain decomposition for
  the initial data solver.  The thick black circles denote the inner
  and outer boundaries of the inner spherical shells (labeled A and B
  next to their centers).  The blue shaded regions represent five open
  cylinders with axis along the line connecting A and B.  The green
  solidly filled regions represent three domains with square
  cross-section.  The thin black circle represents the inner
  boundary of the outer spherical shell, with center indicated by the
  letter C.  Dashed lines are guides to the eye, to indicate the
  dimensions of the various subdomains.  }
\end{figure}

Figure~\ref{fig:Domain} indicates the geometry of the
domain-decomposition employed here.  There are two inner spherical
shells (thick black circles labeled A and B), which are surrounded by
a set of cylinders (light blue) that are aligned with the axis
connecting the two black holes.  

Along the axis of the cylinders there are three subdomains with
rectangular cross-section (indicated in green).  One of these is
located between the two excision spheres, and is a truncated square
pyramid.  The other two are rectangular blocks.  In earlier
work~\cite{Pfeiffer2003} the two inner spherical shells were
restricted to have the \emph{same} outer radius, and all cylinders
were restricted to have the same inner radius.  This restriction
results in a comparatively larger shell around the smaller black hole
(B).  For very unequal mass systems, $m_B\ll m_A$, in particular, it
may be preferable to have a smaller outer radius of shell B, roughly
comparable with the sphere of influence of black hole B.  This would
maximize the agreement of the geometry of the domain decomposition
with the structure of the solution.  Therefore, here, we allow unequal
radii of the two inner shells, as indicated in
Fig.~\ref{fig:Domain}. This has the largest impact when we consider
small separations in initial data (for example, for studying remnant
properties) where the old domain decomposition requires a larger
separation between the two black holes than the new
domain-decomposition in order for the solver to converge.

The new domain-decomposition uses several parameters from which the
placement and dimension of each subdomain follow unambiguously. We
begin by specifying the inner and outer spherical shells:
\begin{itemize}
\item The centres of the inner spherical shells, $\mathbf{c}_{\rm A}$
  and $\mathbf{c}_{\rm B}$, and of the outer spherical shell,
  $\mathbf{c}_{\rm C}$.  Note that $\mathbf{c}_{\rm C}$ is not
  required to lie on the line connecting $\mathbf{c}_{\rm A}$ and
  $\mathbf{c}_{\rm B}$.  
\item The inner and outer radii of the inner spherical shells and the
  outer spherical shell, $r_{\rm A}$, $r_{\rm B}$, $r_{\rm C}$, and $R_{\rm A}$,
  $R_{\rm B}$, $R_{\rm C}$.
\end{itemize}
The remaining parameters $\alpha, f_{\rm cyl}, f_{\rm block},$ and
$f_{\rm C}$ determine the relative sizes of the cylinders and
rectangular blocks:
\begin{itemize}
\item The rectangular blocks and cylinders end on planes orthogonal to
  the axis connecting the centers of the excision spheres. The
  location of these planes is determined by the parameter $\alpha$,
  through the requirement that these planes intersect the inner
  spherical shells A and B in circles of radius $R_{A,B}\sin\alpha$.
  The opening angle of these circles as viewed from the center of the
  spheres is chosen to have the same value for all four planes.

\item The inner radii of the cylinders are determined by the parameter
  $f_{\rm cyl}$ via
\begin{equation}
\rho_{A,B} = f_{\rm cyl} R_{A,B}\sin\alpha.
\end{equation}
Note that $f_{\rm cyl}<1$ is required for the cylinders 1 and 3 to
cover all volume outside the spheres A and B.
\item The size of the blocks orthogonal to the line connecting the two
  spheres is determined by the parameter $f_{\rm block}$,
\begin{equation}
a_{A,B} = f_{\rm block} R_{A,B}\sin\alpha.
\end{equation}
The multiplier $f_{\rm block}$ must satisfy $f_{\rm block}>f_{\rm
  cyl}$ to ensure that the blocks cover the entire open region
within cylinders 0, 2, and 4.
\item The multipler $f_{\rm C}$, which measures how much larger the
  outer size of the cylinders is compared to the inner edge of the
  outer spherical shell:
\begin{equation}
a_{\rm C} = \rho_{\rm C} = f_{\rm C} r_C.
\end{equation}
To ensure complete overlap between the cylinders and the sphere C,
$f_{\rm C}>1+C_\perp / r_{\rm C}$, with $C_\perp$ being the distance
from point $C$ to the axis of the cylinders.
\end{itemize}
The value of $f_{\rm block}$ will determine the relative size of the
face of the blocks to the inner spheres: If $f_{\rm block}>1$, then
the edge of the block will be entirely outside the inner spherical
shell.  Conversely, if $f_{\rm block}<1/\sqrt{2}$, then the face of
the rectangular block is completely contained within the inner
spherical shell.  These considerations will impact which subdomain
(sphere of cylinder) will provide boundary data for the blocks.

Our standard values for the grid-internal geometry coefficients are
$\alpha=45^\circ$, $f_{\rm cyl}=0.95$, $f_{\rm block}=1.05$, and
$f_{\rm C}=1.1$.  We have found these choices to be robust for a wide
variety of component masses, spins and separations.

%%%%%%%%%%%%%%%%%%%%%%%%%%%%%%%%%%%%%%%%%%%%%%%%%%%%%%%%%%%%%%%%
\subsection{Adaptive mesh refinement}
\label{sec:AMR}
%%%%%%%%%%%%%%%%%%%%%%%%%%%%%%%%%%%%%%%%%%%%%%%%%%%%%%%%%%%%%%%%

An important factor in efficiently generating high-accuracy initial
data is the choice of resolution in each of the subdomain used in our
domain decomposition (see Fig.~\ref{fig:Domain}). Typically, we want
our representation of the solution to have about the same accuracy in
all subdomains.  Unfortunately, we do not know a priori what
resolution is needed in a given subdomain to reach a target
accuracy. Furthermore, the optimal resolution varies significantly
with the physical parameters of the binary.  The old initial data
solver~\cite{Pfeiffer2003,Cook2004} used hard-coded resolutions, tuned
to equal-mass low spin BBH.  For unequal mass systems, rapidly
spinning black holes, and/or widely separated binaries the old
resolutions are less efficient and can even prevent convergence of the
elliptic solver when a high accuracy is requested.

To generate initial data, we generally go through multiple
intermediate solves, progressively improving the accuracy of the
solution while converging towards the desired binary parameters. So
instead of predetermining the resolution which will be used in each
subdomain at each level of refinement, we can use the preceeding
numerical solution to predict the optimal resolution in each subdomain
to reach a target accuracy. This significantly improves the efficiency
of the initial data solver, with computing times decreased by about an
order of magnitude for challenging configurations. And it also allows
us to push the binary parameters to more extreme values.

Our multi-domain spectral solver represents the solution in each
subdomain as a tensor-product of basis-functions.  Depending on the
topology of the subdomain, the basis functions are Chebyshev
polynomials, and/or Fourier series, and/or spherical harmonics
(see~\cite{Pfeiffer2003} for details).

Following Szil\'{a}gyi~\cite{Szilagyi:2014fna}, for a given
  subdomain and a given basisfunction, we define the power $P_i$ in
the $i$-th mode by the root-mean-square value of all the coefficients
of the $i$-th mode across all spectral coefficients of the other
basis-functions.  For instance, in a spherical shell with spectral
expansion of the form
$u(r,\theta,\phi)=\sum_{i=0}^{N_r-1}\sum_{l\le L, |m|\le l}\tilde
u_{ilm}T_i(r)Y_{lm}(\theta,\phi)$,
the radial power would be
$P_i=\left(\frac{1}{N_{\theta\phi}}\sum_{l\le L,|m|\le l}|\tilde
  u_{ilm}|^2\right)^{1/2}$,
where $N_{\theta\phi}=(L+1)^2$ represents the number of angular
coefficients\footnote{For spherical harmonic basis-functions, the top
  two modes are filtered~\cite{Pfeiffer2003} and are therefore not
  included in the data $P_i$.}.

For the expected spectral convergence, $P_i$ should decay
exponentially as a function of $i$~\cite{Boyd1999,Pfeiffer2003}, i.e.
$\log_{10}P_i$ when plotted vs. $i$ should be a straight line.  The
slope $f'$ of this line represents the decrease in the magnitude of
the spectral coefficients when going from mode $i$ to mode $i+1$.  We
estimate $f'$ using Eq.~(53) of
Szil\'{a}gyi~\cite{Szilagyi:2014fna}.The current truncation error of
the spectral expansion is approximated as the highest retained
coefficient~\cite{Boyd1999}.

Given the current estimate of the error as $\epsilon$ and the estimate
of the convergence rate as $f'$, we can reach a target accuracy
$\epsilon^*$ by adding
\begin{equation} 
\label{eq:Nupdate}
\Delta N = -\frac{\epsilon-\epsilon^*}{f'}
\end{equation} 
modes to the spectral expansion (recall $f'<0$ and a higher accuracy
means a lower~$\epsilon$).The answer is rounded up so that
$\Delta N > 1$ if the current accuracy is worse than the target
accuracy, and we set $\Delta N = 0$ if $\epsilon<\epsilon^*$, i.e. the
resolution is not allowed to decrease.  For the configuration q3 from
Table~\ref{tbl:Params} the resolution was allowed to decrease without
noticable impact on the convergence behaviour,
cf. Figure~\ref{fig:Many_Convergence}.

The outer spherical shell needs comparatively small angular resolution
$\sim 10$, and sometimes AMR yields the same resolution at neighboring
$E_{DT}$.  Because the ADM-quantities are exclusively evaluated in the
outer spherical shell (cf. Sec.~\ref{sec:AsymptoticQuantities} below),
this would result in apparent non-convergence of ADM linear and
angular momentum.  Therefore, we increase the angular resolution of
the outer sphere by one extra grid-point in the $\theta$ direction and
the corresponding two extra grid-points in the $\phi$ direction,
whenever AMR triggers an adjustment to the domain decomposition.

%%%%%%%%%%%%%%%%%%%%%%%%%%%%%%%%%%%%%%%%%%%%%%%%%%%%%%%%%%%%%%%%%%%%%%%%%%%%%%%%
\subsection{Iterative determination of free parameters}
\label{s:Iterative}
%%%%%%%%%%%%%%%%%%%%%%%%%%%%%%%%%%%%%%%%%%%%%%%%%%%%%%%%%%%%%%%%%%%%%%%%%%%%%%%%
When constructing initial data, we wish to achive desired masses
$M_A^*$, $M_B^*$ and desired black hole spin vectors $\w{\chi}_A^*$
and $\w{\chi}_B^*$. The free data, however, is instead given by the
radii and angular frequencies of the apparent horizons $r_{A,B}$ and
$\mathbf{\Omega}^{H}_{A,B}$, which we write as 
\begin{equation}
\label{eq:u}
\underline{\mathbf{u}}=(r_A,r_B,\mathbf{\Omega}^{H}_{A},\mathbf{\Omega}^{H}_{B}).
\end{equation}
Therefore, one needs to determine values of the free parameters that
result in the desired physical parameters. Thus we must solve the
system of equations
\begin{equation}
  \underline{\mathbf{F}}=(M_{A}-M_{A}^{*},M_{B}-M_{B}^{*},\w{\chi}_{A}-\w{\chi}_{A}^{*},\w{\chi}_{B}-\w{\chi}_{B}^{*})=0.
\end{equation}

The standard approach to the problem would be to use Newton's method;
however, evaluating the Jacobian $J_n$ is too expensive numerically as
every evaluation of the function $\underline{\mathbf{F}}$ requires an elliptic
solve. We instead use the following approach: make an initial
guess $\underline{\mathbf{u}_0}$ based on the Kerr expressions for both black
holes,
\begin{eqnarray}
M_{A,B} &=& r_{A,B}/(1+\sqrt{1-4r^{2}_{A,B}\mathbf{\Omega}_{A,B}^{2}}),\label{eq:Kerr1}\\
\boldsymbol{\chi}_{A,B} &=& - 2r_{A,B}\mathbf{\Omega}_{A,B}^{H},\label{eq:Kerr2}
\end{eqnarray}
and perform an elliptic solve for $\underline{\mathbf{F}}_{0}$. We
then construct an analytic Jacobian $J_{0}$ by using
Eqs.~(\ref{eq:Kerr1},\ref{eq:Kerr2}) to evaluate the partial
derivatives, and update the initial guess by
$\underline{\mathbf{u}}_{1}=\underline{\mathbf{u}}_0-J_{0}^{-1}\underline{\mathbf{F}}_0$.
After this we update the Jacobian using Broyden's
method~\cite{Press2007}:
\begin{equation}
  J_k=J_{k-1}+\frac{1}{\|\underline{\mathbf{\Delta
      u}}_{k}\|}\underline{\mathbf{F}}(\underline{\mathbf{u}}_k)\underline{\mathbf{\Delta u}}_{k}^{T},
\end{equation}
where $\underline{\mathbf{\Delta u}}_{k}= \underline{\mathbf{u}}_{k}-\underline{\mathbf{u}}_{k-1}$. This
corresponds to the ``secant'' approximation for a function of one
variable. Finally we set 
\begin{equation}
  \underline{\mathbf{u}}_{k+1}=\underline{\mathbf{u}}_k-J_{k}^{-1}\underline{\mathbf{F}}_k.
\end{equation}

The major advantage of this approach lies in the use of numerical
information in the update of the Jacobian. This is important in the
regime where the simple analytic Jacobian becomes
inadequate. Broyden's method is applied to the intrinsic physical
properties of each black hole, i.e. the eight parameters listed in
(\ref{eq:u}). We also control more general properties of the binary,
such as the total linear momentum and the position of its centre of
mass. As discussed in Sec.~\ref{subsec:ControlofADM} this is done with
explicit updating formulae that are applied simulateneously at every
step of Broyden's method.

We are now faced with two intertwined iterations: AMR to tune
grid-sizes to a desired truncation error; and root-finding to
adjust free parameters to achieve the desired physical masses, spins,
etc.  When the physical parameters are still far away from the desired
values, very stringent AMR resolution would waste computing time, so
we aim to tighten the AMR resolution while simultaneously decreasing
root-finding errors.  We do so by using an overall truncation error
target $E_{DT}$ for AMR.  We start with a large value for $E_{DT}$,
corresponding to a small grid-size.  As root-finding residuals
decrease, we will decrease $E_{DT}$.  We proceed as follows: At
iteration $k=0,1,2,\ldots$, we compute two measures of progress in
root finding: First, the residual $\mathcal{R}_k$ which quantifies how
close the physical parameters are to their desired values.
$\mathcal{R}_k$ is simply the rms error in the physical parameters:
\begin{equation}
\label{eq:rms_error}
\mathcal{R}_k=\sqrt{\frac{1}{5}\left( \frac{(\Delta M_{A})^{2}+(\Delta M_{B})^{2}}{M^2}+\Delta\w{\chi}_{A}^{2}+\Delta\w{\chi}_{B}^{2}+\frac{\w{P}_{ADM}^{2}}{M^2}\right)}.
\end{equation}
Second, the improvement $\mathcal{I}_k$ that indicates how quickly
root-finding converges, defined as

\begin{equation}
  \mathcal{I}_k=\max_{Q^{i}}\left(\frac{Q^{i}_{k-3}Q^{i}_{k-2}}{Q^{i}_{k-1}Q^{i}_{k}}\right)^{1/2},
      \quad k\ge 3,
\end{equation}
where $Q^{i}=\{\Delta M_{A},\Delta M_{B}, \|\Delta\w{\chi}_A\|,
\|\Delta\w{\chi}_{B}\|,\|\w{P}_{ADM}\|\}$.

We monitor 2 conditions:
\begin{enumerate}
\item $\mathcal{I}_k \le \epsilon_{\mathcal{I}},$
\item $\mathcal{R}_k \le \epsilon_{\mathcal{R}} E_{DT},$ 
\end{enumerate}
where $E_{DT}$ is the desired truncation error, and
$\epsilon_{\mathcal{R}}$ and $\epsilon_{\mathcal{I}}$ are tunable
parameters.  The first condition assures that the resolution is
increased if the root-finding convergence becomes ``flat'' (e.g., due
to the inability to measure the masses accurately enough at the
current resolution). The second condition ensures AMR resolution is
sufficiently high to ensure the physical parameters can be computed
more accurately than the current $\mathcal{R}_k$, with
$\epsilon_{\mathcal{R}}$ being a safety factor.  If either condition
is satisfied and we have already reached our termination truncation
error then the initial data construction is completed.  Otherwise, we
divide $E_{DT}$ by a factor of 10 and continue with the next itertion. For
all cases we have encountered, the choices
$\epsilon_{\mathcal{R}}=10^{2}$ and $\epsilon_{\mathcal{I}}=1.5$ have
proven to be robust.  

%%%%%%%%%%%%%%%%%%%%%%%%%%%%%%%%%%%%%%%%%%%%%%%%%%%%%%%%%%%%%%%%
\subsection{ Calculation of asymptotic quantities}
\label{sec:AsymptoticQuantities}
%%%%%%%%%%%%%%%%%%%%%%%%%%%%%%%%%%%%%%%%%%%%%%%%%%%%%%%%%%%%%%%%

Accurate knowledge of the total energy, linear momentum and angular
momentum of the constructed initial data sets aid their
characterization.  Even more important, accurate control of the total
linear momentum is essential to avoid a drift of the center of mass of
the binary during long evolutions,
cf. Fig.~\ref{fig:PADM_Compare_ToMaster_q3}.

We define the linear and angular momenta on a slice $\Sigma$
intersecting spatial infinity on the surface $S_\infty$ using the
Arnowitt-Deser-Misner (ADM) prescription.  Our initial data satify the
asymptotic gauge conditions~\cite{York:1979}
\begin{eqnarray}
\frac{\partial{\tilde\gamma_{ij}}}{\partial x^j} &=& O(r^{-3}),\\
\gamma^{ij}K_{ij}&=& O(r^{-3}),
\end{eqnarray}
needed to remove ambiguities in the definition of the ADM angular
momentum, as well as the boundary condition $g_{\mu\nu}=\eta_{\mu\nu}$
on $S_\infty$.  The old code~\cite{Pfeiffer2003,Pfeiffer2002a}
directly evaluated the resulting surface integrals at
infinity~\cite{ADM,York:1979},
\begin{eqnarray}
\label{eq:PADM0}
P^i_{\rm ADM} &=& \frac{1}{8\pi}\oint_{S_\infty} (K^{ij}-K\gamma^{ij})\,dS_j,\\
\label{eq:JADM0}
J_i^{\rm ADM} &=& \frac{1}{8\pi}\oint_{S_\infty} \epsilon_{ijk}x^j
(K^{kl}-K\gamma^{kl})\,dS_l, 
\end{eqnarray}
using extrapolation in powers of $1/r$ to infinite
radius~\cite{Pfeiffer2002a}.  $P^{\rm ADM}$ is then found to be a
combination of $1/r^2$ terms of $K^{ij}$, and $J^{\rm ADM}$ a
combination of $1/r^3$ terms.  The old technique, therefore, is very
sensitive to small errors in $K^{ij}$ in the outermost sphere of our
computational domain (the outer boundary is typically located at
$r_{\rm out}\sim 10^{10}M$) and particularly to the presence of
constraint violating modes in that sphere. Typically, this leads to
large errors in $P^{\rm ADM}$ at low resolution, and large errors in
$J^{\rm ADM}$ even at our highest resolution.

Higher accuracy can be obtained by assuming that the constraints are
satisfied on our computational domain, and utilizing Gauss' law to
recast the surface integrals on $S_\infty$ as the sum of a surface
integral on a sphere $S_0$ located at a smaller radius and a volume
integral.  Utilizing $\Psi(S_\infty)=1$, we write
\begin{eqnarray}
 P^i_{\rm ADM}&=& \frac{1}{8\pi}\oint_{S_0}
                  \Psi^{10}(K^{ij}-K\gamma^{ij})\,dS_j\nonumber \\
  && +\frac{1}{8\pi} \int_{V_0}
\frac{\partial}{\partial x^j}\left[\Psi^{10}(K^{ij}-K\gamma^{ij})\right]\,dV.
\end{eqnarray}
Here the normal $dS_j$ to $S_0$ points into the interior of $V_0$
(e.g. along $+\hat{r}$ if it is a coordinate sphere) and the factor
$\Psi^{10}$ was inserted to eliminate terms with spatial derivatives
of $\Psi$ from Eq.~(\ref{eq:G}).  Using the momentum constraint in the
absence of sources,
\begin{equation}
\nabla_j(K^{ij}-\gamma^{ij}K) =
\frac{\partial(K^{ij}-K\gamma^{ij})}{\partial x^j} +
\Gamma^i_{jk}(K^{jk}-\gamma^{jk}K) +
\Gamma^j_{jk}(K^{ik}-\gamma^{ik}K) = 0,
\end{equation}
the volume integral can be simplified to
\begin{equation} \label{eq:PADM}
P^i_{\rm ADM} =
\frac{1}{8\pi}\oint_{S_0} P^{ij}\,dS_j - \frac{1}{8\pi} \int_{V_0}
G^i\,dV.
\end{equation}
Here, 
\begin{eqnarray}
P^{ij} &=& \Psi^{10}(K^{ij}-K\gamma^{ij}),\\
\label{eq:G}
G^i &=& \tilde\Gamma^i_{jk}P^{jk} +
\tilde\Gamma^j_{jk}P^{ik}-2\tilde\gamma_{jk}P^{jk}
\tilde\gamma^{il}\partial_l(\ln\Psi),
\end{eqnarray} 
where $\tilde\Gamma^i_{jk}$ are the connections derived from the
conformal metric $\tilde\gamma_{ij}$. Note that for conformal
flatness and maximal slicing, $G^i=0$ and the volume integral
disappears (see~\cite{FoucartEtAl:2008}).

In practice, for conformally curved initial data, The outer spherical
shell extends to outer radius $\sim 10^{10}M$.  Therefore, in the
numerical evaluation of the volume integral in Eq.~(\ref{eq:PADM}),
the volume element associated with the outermost grid-point becomes
very large and introduces numerical noise.  To avoid this, we roll off
the integrand $G^{i}$ beyond a certain radius $R_c$, i.e.  we replace
$G^i$ by $\tilde G^i$ given by

\begin{equation}
  \tilde{G}^i =
  \left\{
        \begin{array}{ll}
            G^i, & \quad r \leq R_c, \\
            \frac{R_c^2}{r^2} G^i,& \quad r > R_c.
        \end{array}
    \right.
\end{equation}
We choose $R_c=1000\max(w_A, w_B)$, where $w_{A,B}$ are the widths of the Gaussians that enforce exponential falloff to conformal flatness (cf.~Eqs.~45 and 46 of Lovelace et al~\cite{Lovelace2008}).

The ADM angular momentum is also rewritten using Gauss' law as %
\begin{equation} J^z_{\rm ADM} =
\frac{1}{8\pi}\oint_{S_0} (x P^{yj}-yP^{xj})\,dS_j - \frac{1}{8\pi}
\int_{V_0}(xG^y-yG^x)\,dV,
\label{eq:Jsv}
\end{equation}
with cyclical permutations of $(x,y,z)$ yielding the other components.
For maximal slicing and conformal flatness in $V_0$,
Eq.~(\ref{eq:Jsv}) simplifies to
\begin{equation}
J^z_{\rm ADM} = \frac{1}{8\pi}\oint_{S_0} \Psi^{10}(xK^{yj}-yK^{xj})\,dS_j.
\label{eq:Js}
\end{equation}
Because Eq.~(\ref{eq:Jsv}) relies on the cancellation of large volume
terms, it can be sensitive to errors in $K^{ij}$ .  Accordingly, we
use Eq.~(\ref{eq:Js}) using a surface $S_0$ at sufficiently large
radius such that in $V_0$ the metric is conformally flat and $K=0$.

\begin{figure}
\centering
  \includegraphics[width=0.7\linewidth]{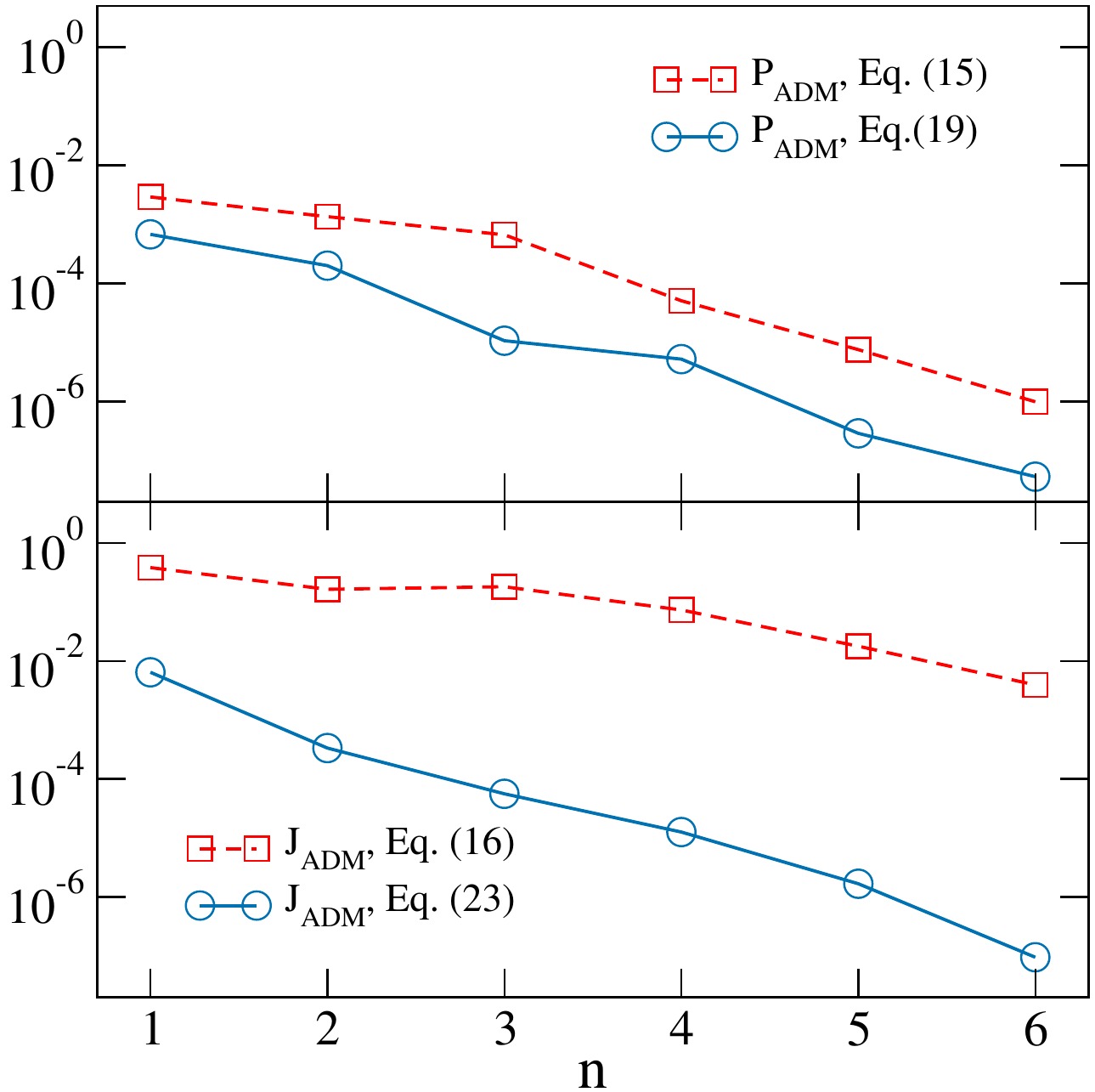}
  \caption{\label{fig:JADM_convergence_q50} Accuracy of the
    calculation of $P_{\rm ADM}$ and $J_{\rm ADM}$ for two different
    methods of evaluation.  We evaluate $P_{\rm ADM}$ and
    $J_{\rm ADM}$ when truncation error $E_{DT}=10^{-n-3}$ is reached,
    and plot differences to the next lower resolution $n-1$.  Data
    shown for case q50 in Table~\ref{tbl:Params}. }
\end{figure}

To illustrate the importance of the transformations applied to the ADM
integrals, we consider the convergence test for configuration q50.  We
evaluate $P_{\rm ADM}$ using Eq.~(\ref{eq:PADM0}) and
Eq.~(\ref{eq:PADM}), and we evaluate $J_{\rm ADM}$ using
Eq.~(\ref{eq:JADM0}) and Eq.~(\ref{eq:Jsv}).
Figure~\ref{fig:JADM_convergence_q50} shows the results.

The calculation of $P_{\rm ADM}$ is improved by about one order of
magnitude when utilizing Gauss' law, whereas $J_{\rm ADM}$ improves by
several orders of magnitude.  We point out that, in order to achieve
\emph{any} convergence for the old $J_{\rm ADM}$ calculation, we had
to manually increase the radial resolution in the outer sphere by $1$
whenever the domain decomposition is adjusted.

We also compute a new diagnostic, the centre-of-mass
$\mathbf{C}_{\rm CoM}$ of the initial data sets using the formalism
developed in Ref.~\cite{Baskaran:2003pk}.  In conformal flatness, the
expressions from~\cite{Baskaran:2003pk} reduce to
\begin{equation}\label{eq:CoM}
\mathbf{C}_{\rm CoM} = \frac{3}{8\pi\,  E_{\rm ADM}}\;\lim_{R\to\infty}\oint \Psi^4\,\mathbf{n}\, dA,
\end{equation}
where $\mathbf{n}$ is the outward-pointing unit normal, $\mathbf{n}=\mathbf{r}/r$.
Equation~(\ref{eq:CoM}) is numerically evaluated by expanding the
conformal factor $\Psi$ in a power-series in $1/r$.  We read off the
(angle-dependent) coefficient of the $1/r^2$ term, and expand this
coefficient in spherical harmonics.  Each individual spherical
harmonic term can be integrated against $\mathbf{n}$ analytically, so
that the integral~(\ref{eq:CoM}) collapses to a linear combination of
spherical-harmonic coefficients.

%%%%%%%%%%%%%%%%%%%%%%%%%%%%%%%%%%%%%%%%%%%%%%%%%%%%%%%%%%%%%%%%
\subsection{Control of linear momentum and centre of mass}
\label{subsec:ControlofADM}
%%%%%%%%%%%%%%%%%%%%%%%%%%%%%%%%%%%%%%%%%%%%%%%%%%%%%%%%%%%%%%%%

The quasi-equilibrium conformal thin-sandwich formalism to construct
binary black hole initial data was developed in a series of
papers~\cite{Cook2004,Caudill-etal:2006,Pfeiffer-Brown-etal:2007,Lovelace2008,Buchman:2012dw}.
In this formalism, one chooses two excised regions (usually taken to
be coordinate spheres) with centres $\mathbf{c}_{A,B}$, and solves the
extended conformal thin sandwich
equations~\cite{York1999,Pfeiffer2003b} in the exterior.  Boundary
conditions on the excised regions ensure that they are apparent
horizons, and control the spin of each black hole.  The locations and
the sizes of the excised regions correlate with the position and
masses of the two black holes.  Orbital rotation is induced by the
requirement that certain time-derivatives vanish in a frame rotating
with orbital velocity $\mathbf{\Omega}_0$ about the orign.  One
finally incorporates a radial expansion factor $\dot{a}_0$, which
allows fine control of the orbital
eccentricity~\cite{Pfeiffer-Brown-etal:2007,Boyle2007,Buonanno:2010yk,Buchman:2012dw}.
By a suitable choice of the conformal quantities, the
quasi-equilibrium approach can generate initial data with black hole
spins of order $0.9998$~\cite{Lovelace2008}.

One shortcoming of the formalism presented in~\cite{Buchman:2012dw}
lies in a lack of control of the center of mass of the binary, and
only incomplete control of the ADM linear momentum
$\mathbf{P}_{\rm ADM}$.  The past implementations use the location of
the black holes to partially control $\mathbf{P}_{\rm ADM}$.  Consider
a small displacement $\delta\mathbf{c}$ applied to the centres of both
excision regions.  Through the orbital rotation $\mathbf{\Omega}_0$
about the origin, the displacement $\delta\mathbf{c}$ induces a change
in velocity of the black holes of
$\mathbf{\Omega}_0\times\delta\mathbf{c}$, with a corresponding change
in $\mathbf{P}_{\rm ADM}$.  Therefore, $\delta\mathbf{c}$ could be
used to cancel the components of $\mathbf{P}_{\rm ADM}$ orthogonal to
$\mathbf{\Omega}_0$; however, the cross-product in
$\mathbf{\Omega}_0\times\delta\mathbf{c}$ prevented any correction
parallel to $\mathbf{\Omega}_0$.  For head-on collisions with
$\mathbf{\Omega}_0=0$, no control of $\mathbf{P}_{\rm ADM}$ is possible
at all.  For the non-precessing simulations presented
in~\cite{Buchman:2012dw}, the component of $\mathbf{P}_{\rm ADM}$
parallel to $\mathbf{\Omega}_0$ vanishes by symmetry, and no problems
arose.  However, for generic precessing binaries, there will be a
non-zero linear momentum orthogonal to the orbital plane, which
results in a drift of the center of mass for very long simulations
(see~\cite{Ossokine:2013zga} for an extreme example).

Here, we propose a different means to control the full
$\mathbf{P}_{\rm ADM}$, while simultaneously allowing us to control
the center of mass as well.  We fix the relative separation of the
centres of the excision spheres,
\begin{equation}
\mathbf{c}_A-\mathbf{c}_B = \mathbf{D},
\end{equation}
where the separation vector $\mathbf{D}$ is user-specified.  We use
the choice of $\mathbf{c}_A$ to control the center-of-mass
$\mathbf{C}_{\rm CoM}$ of the binary.  Once a first initial data set is
computed (with, in general, $\mathbf{C}_{\rm CoM}\neq \mathbf{0}$), we
can update 
\begin{equation}\label{eq:cA-update}
   \mathbf{c}_{{\rm A},k+1} = \mathbf{c}_{{\rm A},k} -
   \mathbf{C}_{{\rm CoM}, k} - \frac{M_{{\rm A},k}\Delta M_{{\rm B}, k}-M_{{\rm B},k}\Delta
   M_{{\rm A},k}}{(M_{{\rm A},k}+M_{{\rm B},k})^{2}}\,\mathbf{D}.
\end{equation}

With the black-hole centres now used to control the centre of mass, we
need a different means to control $\mathbf{P}_{\rm ADM}$.  We add in
the outer boundary condition on the shift (Eq. (38c)
of~\cite{Lovelace2008}) a constant velocity $\mathbf{v}_0$:
\begin{equation}\label{eq:ShiftBC}
\beta^i = (\mathbf{\Omega}_0\times\mathbf{r})^i+\dot{a}_0r^i +v_0^i
\qquad\mbox{on $\,\mathcal B$}.
\end{equation}
Here $\mathcal B$ represents the outer boundary, a sphere with radius
$R=10^{10}M$.  The velocity $\mathbf{v}_0$ will effect the overall
motion of the binary, and will be reflected in a corresponding change
in $\mathbf{P}_{\rm ADM}$ by $E_{\rm ADM} \mathbf{v}_0$, where
$E_{\rm ADM}$ is the ADM-energy of the binary.  During iterative
root-finding of the free parameters, we adjust $\mathbf{v}_0$ to
achieve $\mathbf{P}_{\rm ADM}=\mathbf{0}$\footnote{Using the obtained
  vector $\beta^i$ as the shift-vector in an \emph{evolution} results
  in a translating outer boundary; this effect is eliminated by
  evolving with a shift vector of $\beta^i-v_0^i$.  }.

\begin{figure}
\centering
  \includegraphics[width=0.7\linewidth]{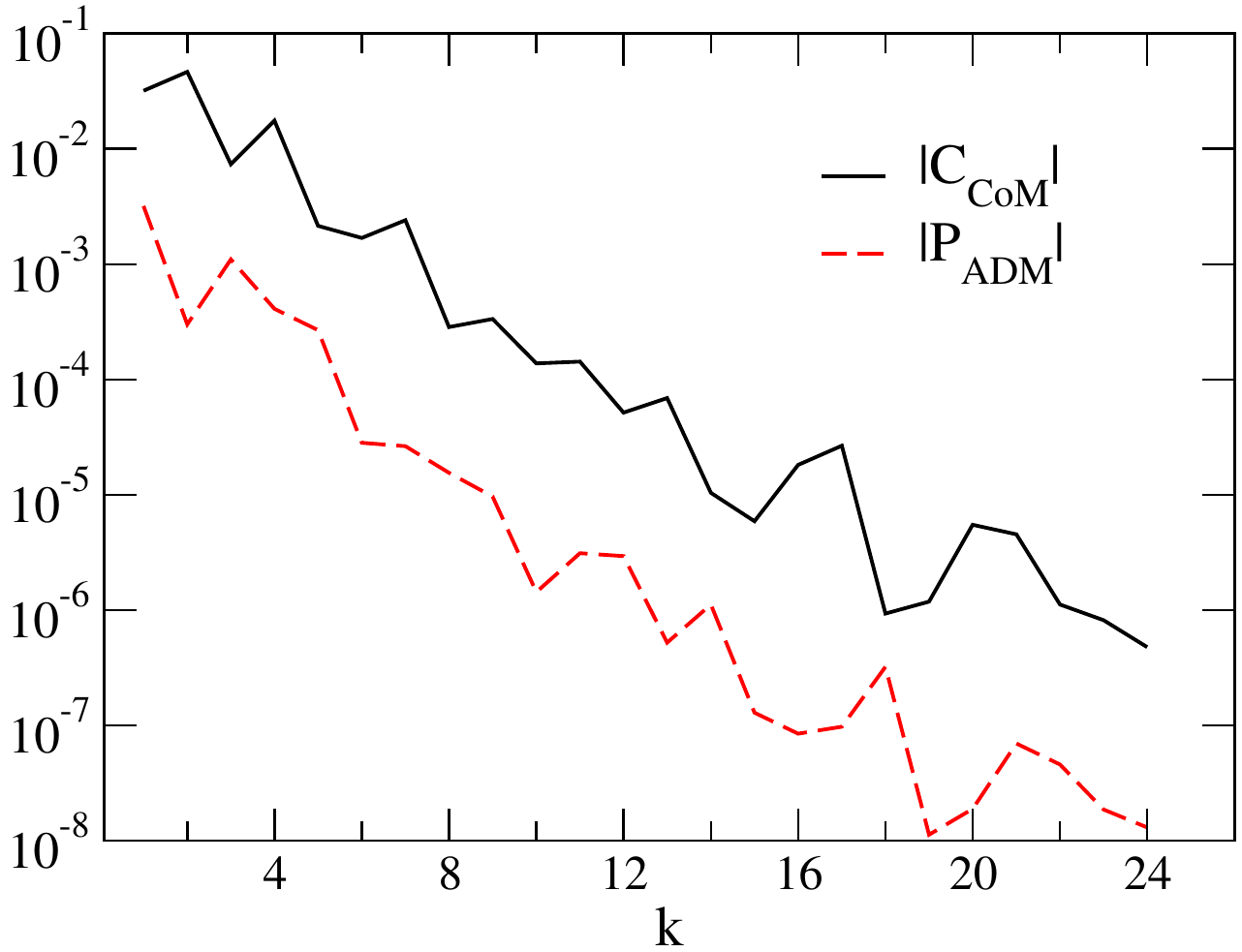}
  \caption{\label{fig:Convergence_PADM} Control of the centre
      of mass and linear momentum for a $q=10$ generically precessing
      binary (see case q10 in Table~\ref{tbl:Params}) . Shown are the
      magnitude of $\w{C}_{\rm CoM}$ (black solid lines) and
      $\mathbf{P}_{\rm {ADM}}$ (red dashed lines) as a function of
      root-finding iteration $k$.}
\end{figure}

To motivate the updating formula for $\mathbf{v}_0$, consider a
perturbation of $\mathbf{v}_0$ by $\delta\mathbf{v}_0$, and a
perturbation of $\mathbf{c}_A$ by $\delta \mathbf{c}$. 
If we allow the masses to vary, then a Newtonian-inspired formula is
 \begin{equation}\label{eq:v0-update}
\mathbf{v}_{0,k+1} = \mathbf{v}_{0,k} - \frac{\mathbf{P}_{{\rm ADM},k}}{M_k} + (\Delta M_{{\rm A},k}+\Delta
M_{{\rm B}, k})(\mathbf{v}_k+\mathbf{\Omega}\times\mathbf{c}_{{\rm A},k}) - \mathbf{\Omega}\times
\mathbf{\delta c}_{{\rm A}, k} - \frac{\Delta M_{{\rm B}, k}}{M_k} \mathbf{\Omega}\times
\mathbf{D}.
\end{equation}

To summarize, relative to earlier initial-data sets, we modify the
outer boundary condition for the shift by the term $\mathbf{v}_0$,
cf. Eq.~(\ref{eq:ShiftBC}), and use updating
formulae~(\ref{eq:cA-update}) and~(\ref{eq:v0-update}) to adjust
$\mathbf{c}_A$ and $\mathbf{v}_0$.
Section~\ref{sec:AsymptoticQuantities} describes how we compute
$\mathbf{P}_{\rm ADM}$ and $\mathbf{C}_{\rm CoM}$.

We demonstrate the efficiency of the updating formulas
Eqs.~(\ref{eq:cA-update},\ref{eq:v0-update}) in
Fig. \ref{fig:Convergence_PADM} that shows the magnitude of
$\mathbf{C}_{\rm CoM}$ and $\mathbf{P}_{\rm ADM}$ as a function of
root-finding iteration for a $q=10$ precessing binary (case q10 in
Table~\ref{tbl:Params}). The convergence is evidently very fast, with
the final values of $\sim 10^{-6}$ and $\sim 10^{-8}$
respectively. This means that even for an inspiral lasting $10^{6}$ M,
the drift of the centre of mass due to residual linear momentum in
initial data will be only $\sim 0.01M$.

%%%%%%%%%%%%%%%%%%%%%%%%%%%%%%%%%%%%%%%%%%%%%%%%%%%%%%%%%%%%%%%%
\section{Numerical results}
\label{sec:NumericalResults}
%%%%%%%%%%%%%%%%%%%%%%%%%%%%%%%%%%%%%%%%%%%%%%%%%%%%%%%%%%%%%%%%
\subsection{Initial data construction}
\label{s:IDconstr}
We test the improvements described in the previous sections on several
cases of interest, whose parameters are summarized in
Table~\ref{tbl:Params}. The parameters were chosen to demonstrate the
range of initial data sets that can be constructed with the new code
and to provide some overlap with regions of parameter space which
could be achieved previously.

\begin{table}
  \caption{  \label{tbl:Params}
    Physical parameters of the initial data sets used for testing the new initial data code.} 
\centerline{
  \begin{tabular}{c|ccc|ll}
        Name & q & $\w{\chi}_{1}$ & $\w{\chi}_{2}$ & $D_0/M$ &
                                                           $M\Omega_{0}$ \\
     \hline
    Spin0.9999 & 1 & (0, 0, 0.9999) & (0, 0, 0.9999) & 14.17 & 0.01682 \\
    q3 & 3 & (0, 0.49, -0.755) & (0, 0, 0) & 15.48 & 0.01515 \\ 
    q10 & 10 & (0.815, -0.203,  0.525) & (-0.087, 0.619, 0.647)  &
                                                       15.09 &  0.01547\\
    q50 & 50 &
               (-0.045, 0.646, -0.695)
                              & (0, 0, 0) &  16  & 0.01428 \\
    \end{tabular}
}
\end{table}

\begin{figure}
\centering
\includegraphics[width=0.72\linewidth]{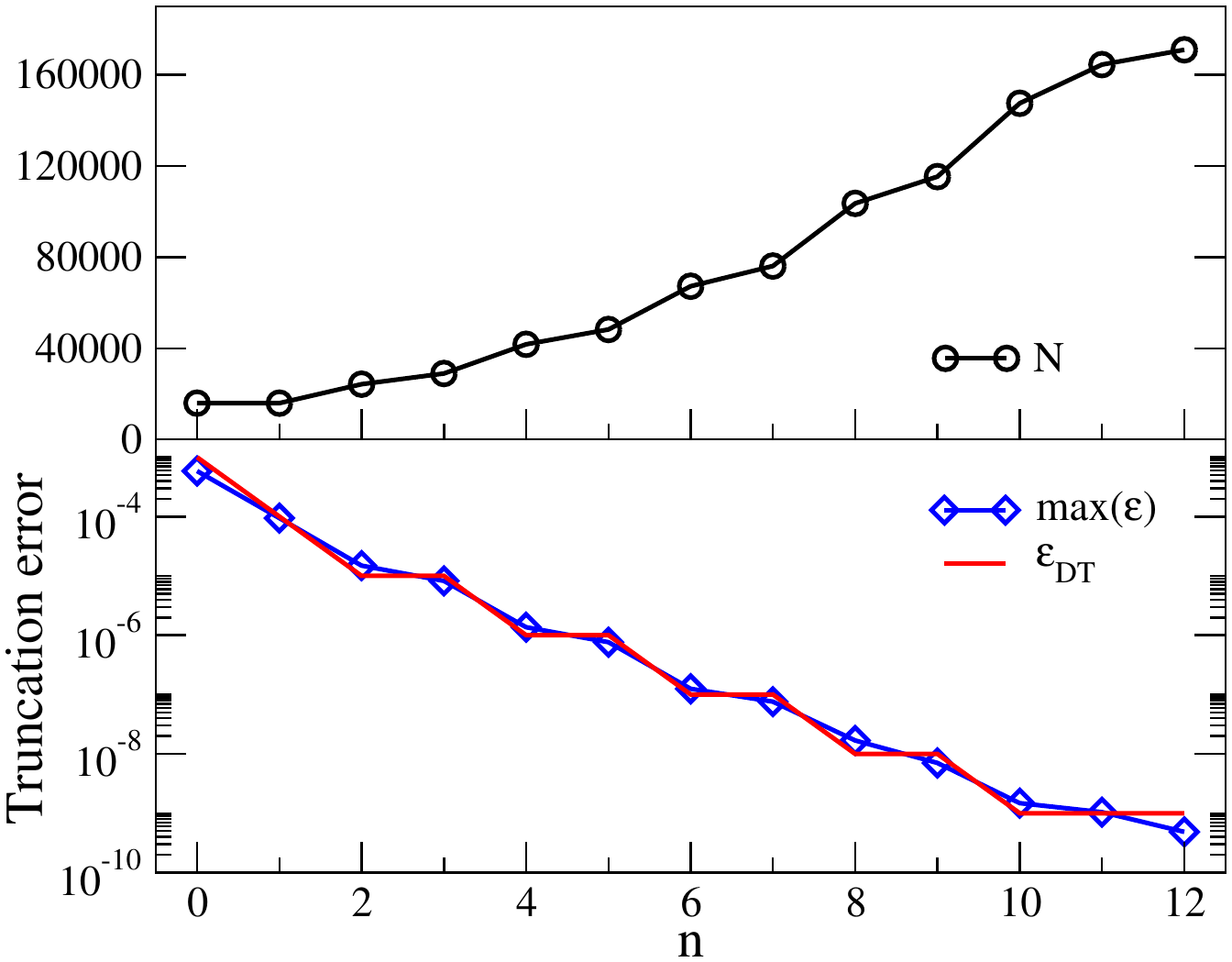}
\caption{\label{fig:AMR_work_q3} Behaviour of the AMR algorithm for
  case q3  as function of AMR iteration $n$. {\bf Top}: total
  number of collocation points. {\bf Bottom}: Highest truncation error
  in all subdomains, and target truncation error. The free parameters
  are fixed to their values at the end of root-finding.}
\end{figure}

We first illustrate the performance of the AMR outlined in
Sec.~\ref{sec:AMR} with the case q3, a configuration we will compare
with the old BBH solver below. To demonstrate AMR in isolation, we fix
initial data parameters, and start with target truncation error
$E_{DT}=10^{-3}$.  We solve the constraint equations, estimate
spectral truncation errors and update numerical resolution via
Eq.~(\ref{eq:Nupdate}).  Whenever we reach the desired truncation
error, we tighten the AMR error tolerances by dividing $E_{DT}$ by 10,
until a truncation error of $10^{-9}$ is reached.
Figure~\ref{fig:AMR_work_q3} illustrates the behaviour of the AMR
algorithm during this test. The top panel shows the total number of
collocation points in the domain, which grows with each AMR
iteration. The bottom panel demonstrates that the largest truncation
error across all subdomains, $\max\epsilon$, closely tracks the
truncation error target $E_{DT}$.

\begin{figure}
\begin{center}
  \includegraphics[width=0.8\linewidth]{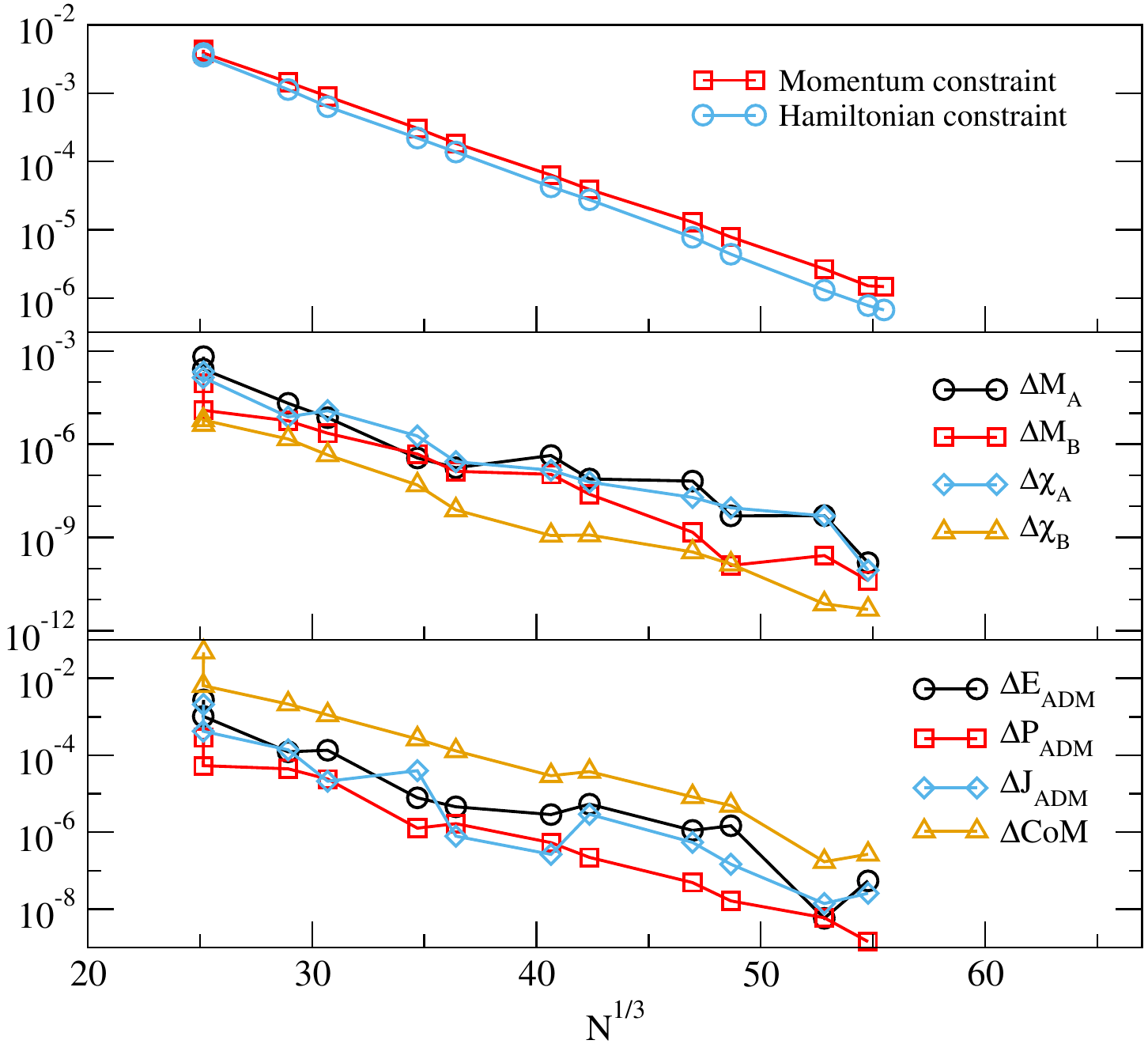}
  \caption{ \label{fig:Convergence_q3} Convergence of physical
    quantities with increasing resolution for case q3. {\bf Top}: the
    norm of Hamiltonian and momentum constraints, {\bf middle}: masses
    and spins, {\bf bottom}: ADM quantities and center of mass.
    ``$\Delta$'' indicates the difference between the value of the
    respective quantity at the current resolution and the highest
    resolutions. The free parameters are fixed to their values at the
    end of root finding.}
\end{center}
\end{figure}

Figure ~\ref{fig:Convergence_q3} shows a convergence test of the AMR
sequence shown in Fig.~\ref{fig:AMR_work_q3}.  Plotted are various
quantities as a function of the effective number of grid-points
$N^{1/3}$. The top panel demonstrates the exponential decrease in the
$L^{2}$ norms of the Hamiltonian and momentum constraints, which
implies that this data set is constraint-satisfying. The constraints
are given explicitly by:
\begin{eqnarray}
C_{\rm Ham}&=&\frac{1}{2}\left(R+K^{2}-K_{ab}K^{ab}\right), \\
C_{\rm Mom}&=&D_{b}K^{b}_{a}-D_{a}K,
\end{eqnarray}
where $D$ is the covariant derivative associated with the spatial
metric. The $L^{2}$ norm is simply the normalized pointwise norm over all
collocation points:
\begin{equation}
\|s\|_{L^{2}} = \sqrt{\frac{1}{N}\sum_{i}^{N}s_{i}^{2}}.
\end{equation}

The convergence of the masses and spins is shown in the
middle panel. Here we plot the norms of the differences between the
quantity at a given iteration and its value at the highest resolution:
\begin{equation}
\Delta Q = |Q_{i}-Q_{\rm max}|. \qquad
\end{equation}
Once again, the convergence is essentially exponential. The bottom
panel of Fig.~\ref{fig:Convergence_q3}, finally examines the
convergence of the ADM quantities and the center-of-mass computation.
Though convergence is not as clean as for the constraints, the bottom
panel of Figure~\ref{fig:Convergence_q3} shows that all the asymptotic
quantities can be determined to better than $10^{-6}$.

\begin{figure}
  \begin{center}
    \includegraphics[width=0.7\linewidth]{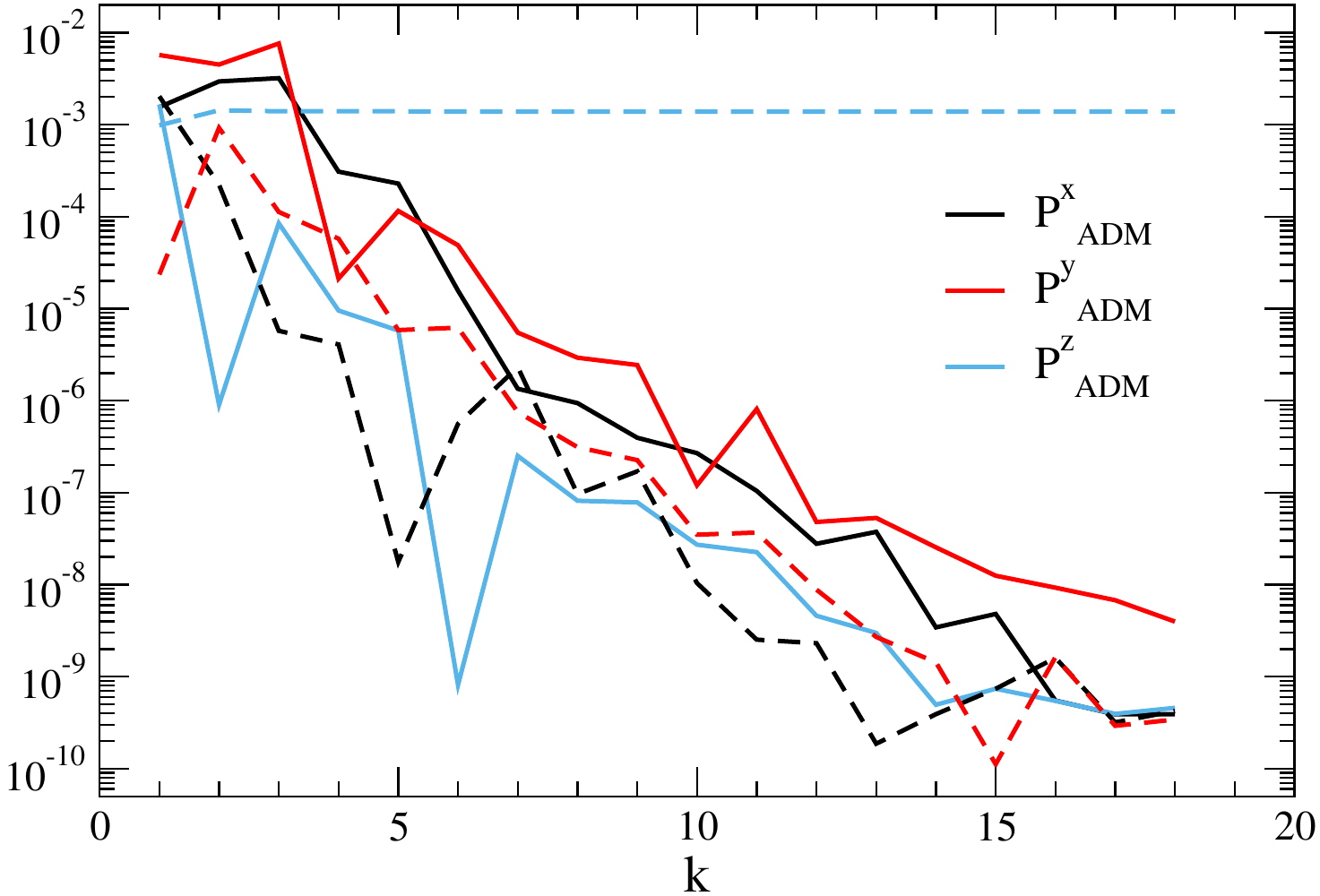}
  \end{center}
  \caption{\label{fig:PADM_Compare_ToMaster_q3} Comparison of
    $\mathbf{P}_{\rm ADM}$ control between the new code (solid lines)
    and the old code (dashed lines) as a function of root-finding
    iteration k. Both versions successfully control the $x$ and $y$
    components of the linear momentum, but only the new code controls
    the z-component as well.}
\end{figure}

To conclude our detailed examination of the initial data set {\tt q3},
we contrast the new code described here with the old
code~\cite{Pfeiffer2003,Buchman:2012dw}. One of the most important
upgrades lies in the control of $\w{P}_{\rm ADM}$. Figure
~\ref{fig:PADM_Compare_ToMaster_q3} shows the components of
$\w{P}_{\rm ADM}$ as a function of root-finding iteration $k$ for both
the new and the old code.\footnote{Both codes \emph{compute}
  $\w{P}_{\rm ADM}$ in the same way (via \cref{eq:PADM}), but differ
  in the way it is \emph{controlled}.} Both codes successfully drive
$P_{\rm ADM}^x$ and $P_{\rm ADM}^y$ to zero as expected.  But only the
new code also drives $P_{\rm ADM}^z$ to zero, whereas the old code
yields $P_{ADM}^{z}\simeq 0.00138$ . As we shall see in the next
section, this produces non-trivial differences in the evolution.

\begin{figure}

  \begin{center}
    \includegraphics[width=0.78\linewidth]{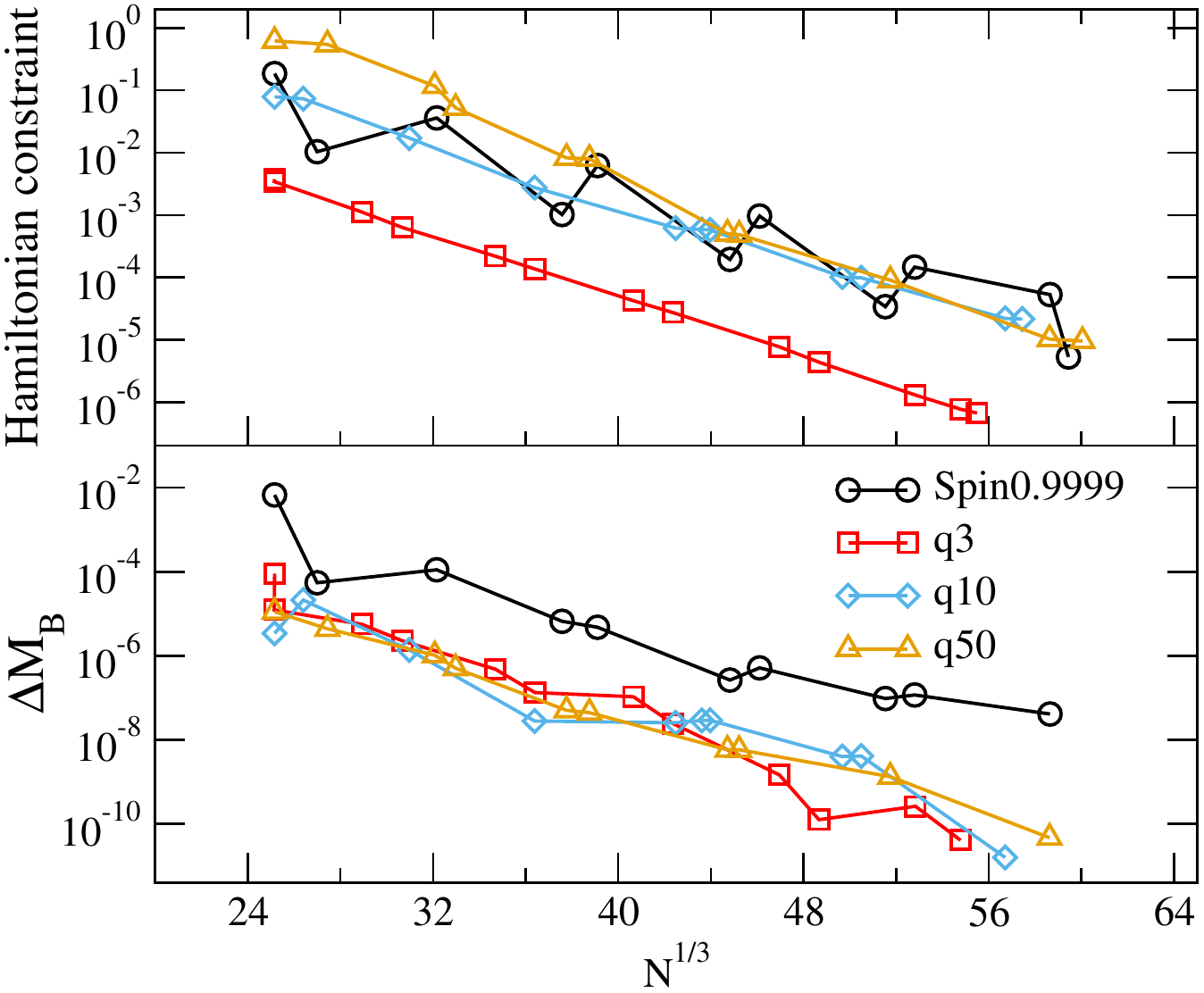}
  \end{center}
  \caption{ \label{fig:Many_Convergence} Overview of initial data
    results for cases in Table~\ref{tbl:Params}. {\bf Top}:
    convergence of the $L^{2}$ norm of the Hamiltonian
    constraint. {\bf Bottom}: convergence of the mass of the
    \emph{smaller} black hole.The free parameters are
    fixed to their values at the end of root-finding. }
\end{figure}

Turning to the more challenging cases listed in
Table~\ref{tbl:Params}, we have performed similar tests to case {\tt
  q3}, with the free parameters fixed to their values at the end of
root finding and only the resolution changing from iteration to
iteration.  As an example, Figure~\ref{fig:Many_Convergence} shows a
subset of the convergence data. This figure demonstrates that the
exponential convergence shown previously for case {\tt q3} extends to
all cases. In particular, the constraints are exponentially
convergent.  All four cases complete with a maximal resolution of less
than $2.5\times10^{5}$ points, an improvement of a factor of $2-3$
over the old code.\footnote{We note that the case q50 could not be
  constructed with the old code.} Owing to the more challenging
configurations, however, the constraints are 1-2 orders of magnitude
larger. The physical parameters are also exponentially converegent
with resolution, as illustrated in the lower panel of
Fig.~\ref{fig:Many_Convergence}. We use $\Delta M_{B}$ since it is
frequently harder in a high mass ratio to resolve the smaller black
hole, so this provides a conservative convergence test.

%%%%%%%%%%%%%%%%%%%%%%%%%%%%%%%%%%%%%%%%%%%%%%%%%%%%%%%%%%%%%%%%
\subsection{Root-finding}
%%%%%%%%%%%%%%%%%%%%%%%%%%%%%%%%%%%%%%%%%%%%%%%%%%%%%%%%%%%%%%%%

\begin{figure}
  \begin{center}
    \includegraphics[width=0.95\linewidth]{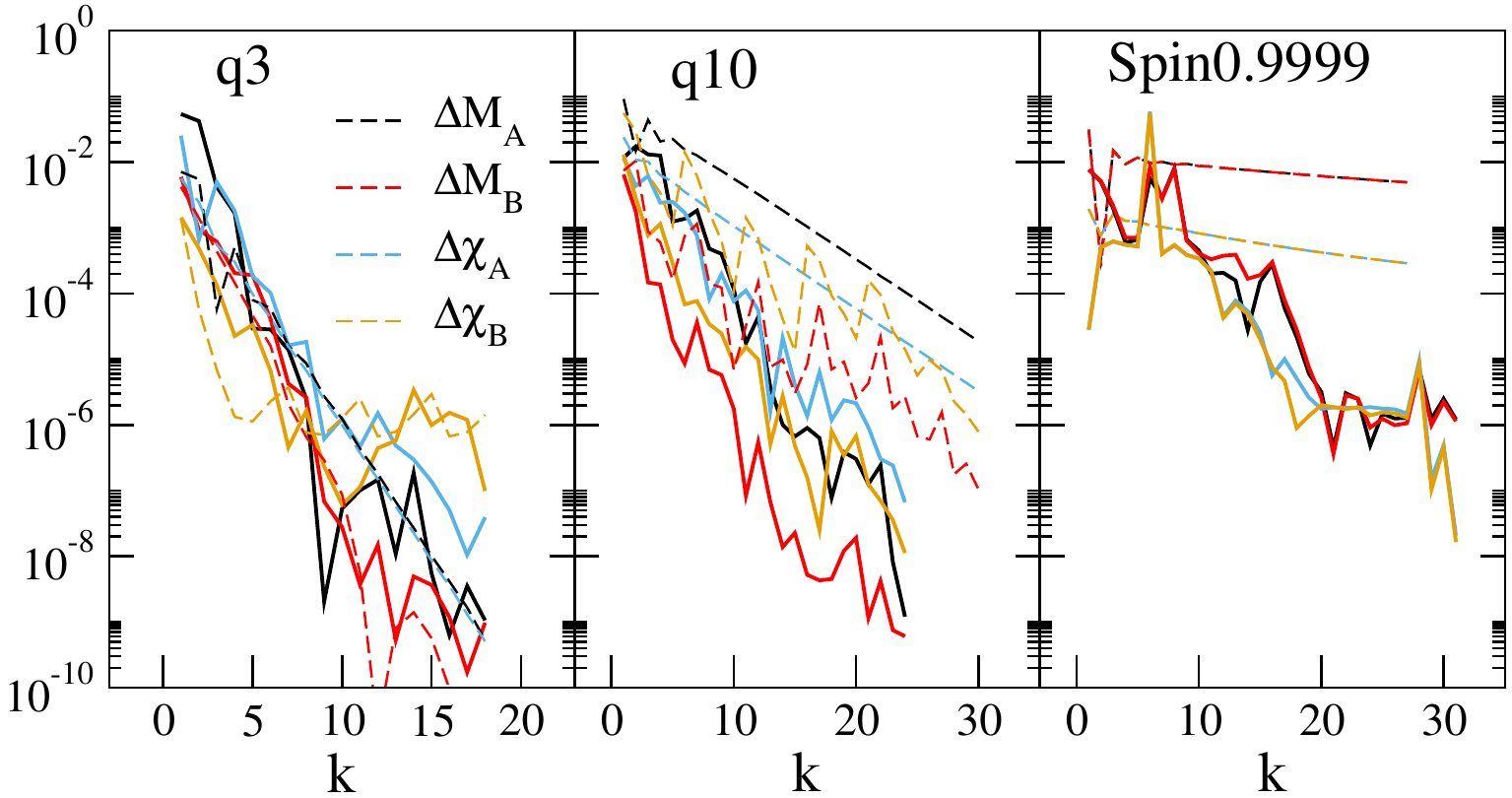}
  \end{center}
  \caption{\label{fig:RootFinding} Convergence of the root-finding
    procedure for masses and spins when the old updating formulae are
    used (dashed lines), and with the new updating formulae developed
    here (solid lines).  For q3 both algorithms perform well, whereas
    for q10 the new code converges about twice as fast as the old
    code. Finally for Spin0.9999 the old code fails to achieve desired
    masses and spins, while the new code gives errors of order
    $10^{-6}$. }
\end{figure}

It is also important to examine the performance of the updated
root-finding procedure based on Broyden's
method. Figure~\ref{fig:RootFinding} shows the root-finding results
for cases q3, q10 and Spin0.9999 done with the old and new versions of the
code. Note that during root-finding the resolution of the subdomains
is also allowed to change to achieve the desired truncation error. For
low mass ratio both codes show similar rates of convergence and final
errors. The situation changes for case q10, where the old code has
trouble achieving low errors in masses and spins, while the new
root-finding procedure described in Sec.~\ref{s:Iterative} results in
errors of order $10^{-6}$. Finally, for case Spin0.9999, the results are
drastically different: the old code has errors in the masses of order
several $\times 10^{-3}$ and spins of order $10^{-4}$. Since we are
attempting to construct a binary with dimensionless spins of $0.9999$
it becomes clear that the old code is inadequate for this purpose. On
the other hand, the new root-finding procedure successfully reduces
the errors in physical quantities to the level of $10^{-6}$.  Thus,
the new algorithm allows us to achieve the desired values of the
physical quantities which is especially important as we push to higher
spin magnitudes.

On average, the new code is about 25-50\% as fast as the old
one.  For example, for the case q10, the old code took 12.4 hours to
complete, whereas the new took 6 hours on 12 cores of a Westmere node
of the Briar\'{e}e compute cluster. Therefore, the new code is indeed more
efficient than the old while achieving the same or better accuracy.

%%%%%%%%%%%%%%%%%%%%%%%%%%%%%%%%%%%%%%%%%%%%%%%%%%%%%%%%%%%%%%%%
\subsection{Exploratory evolution}
\label{sec:Evolutions}
%%%%%%%%%%%%%%%%%%%%%%%%%%%%%%%%%%%%%%%%%%%%%%%%%%%%%%%%%%%%%%%%

\begin{figure}
  \begin{center}
    \includegraphics[width=0.8\linewidth]{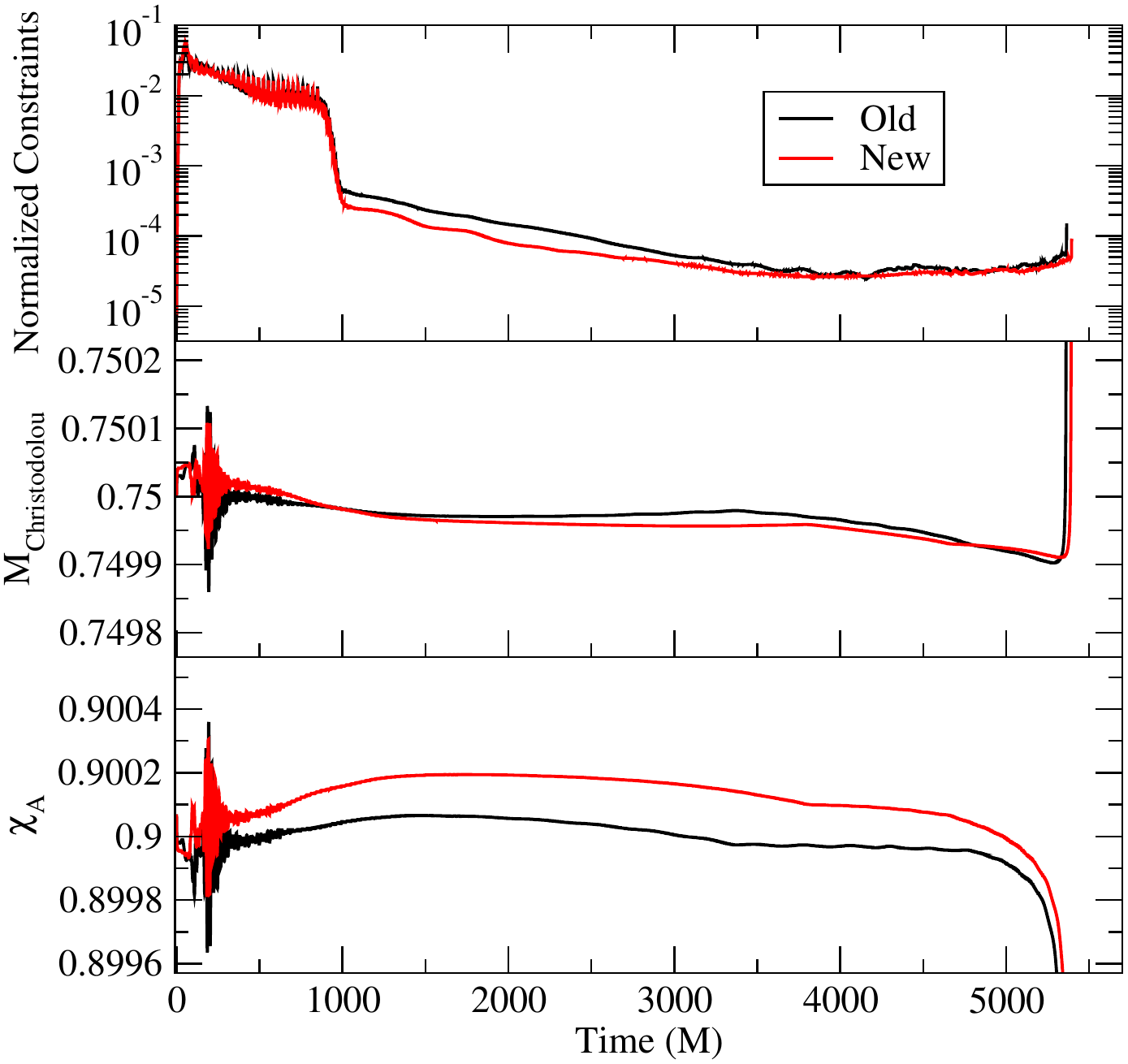}
  \end{center}
  \caption{\label{fig:q3_EvolutionConvergence} {\bf Top}: Evolution of
    the normalized constraints. {\bf Middle}: Evolution of the
    Christodolou mass of the large black hole. {\bf Bottom}: Evolution
    of the spin $|\w{\chi}_{A}|$.}
\end{figure}

We have emphasized above the importance of controlling
$\w{P}_{\rm ADM}$.  We now evolve initial data for case {\tt q3}
constructed with the old and the new initial-data code, and compare
the two evolutions in detail.

We being by considering the convergence of constraints and quasi-local
quantities during the evolution. The top panel of
Figure~\ref{fig:q3_EvolutionConvergence} shows the $L^{2}$ norm of the
normalized constraint violations during the evolution (see Eq.(71)
of~\cite{Lindblom2006}). It is obvious that both codes show similar
convergence properties, as expected. Further, the initial spike of
constraint violations due to junk radiation is virtually
indistinguishable, which indicates that the new method of constructing
initial data does not introduce additional constraint-violating modes.
The middle and bottom panels of Fig.~\ref{fig:q3_EvolutionConvergence}
show the evolution of the Christodoulou mass and the spin magnitude of
the large black hole.  The differences between the evolutions of the
old and new initial data sets are consistent with truncation
error. Thus we conclude that the quasi-local quantities are the same
in both data sets.

\begin{figure}
  \begin{center}
    \includegraphics[width=0.49\linewidth]{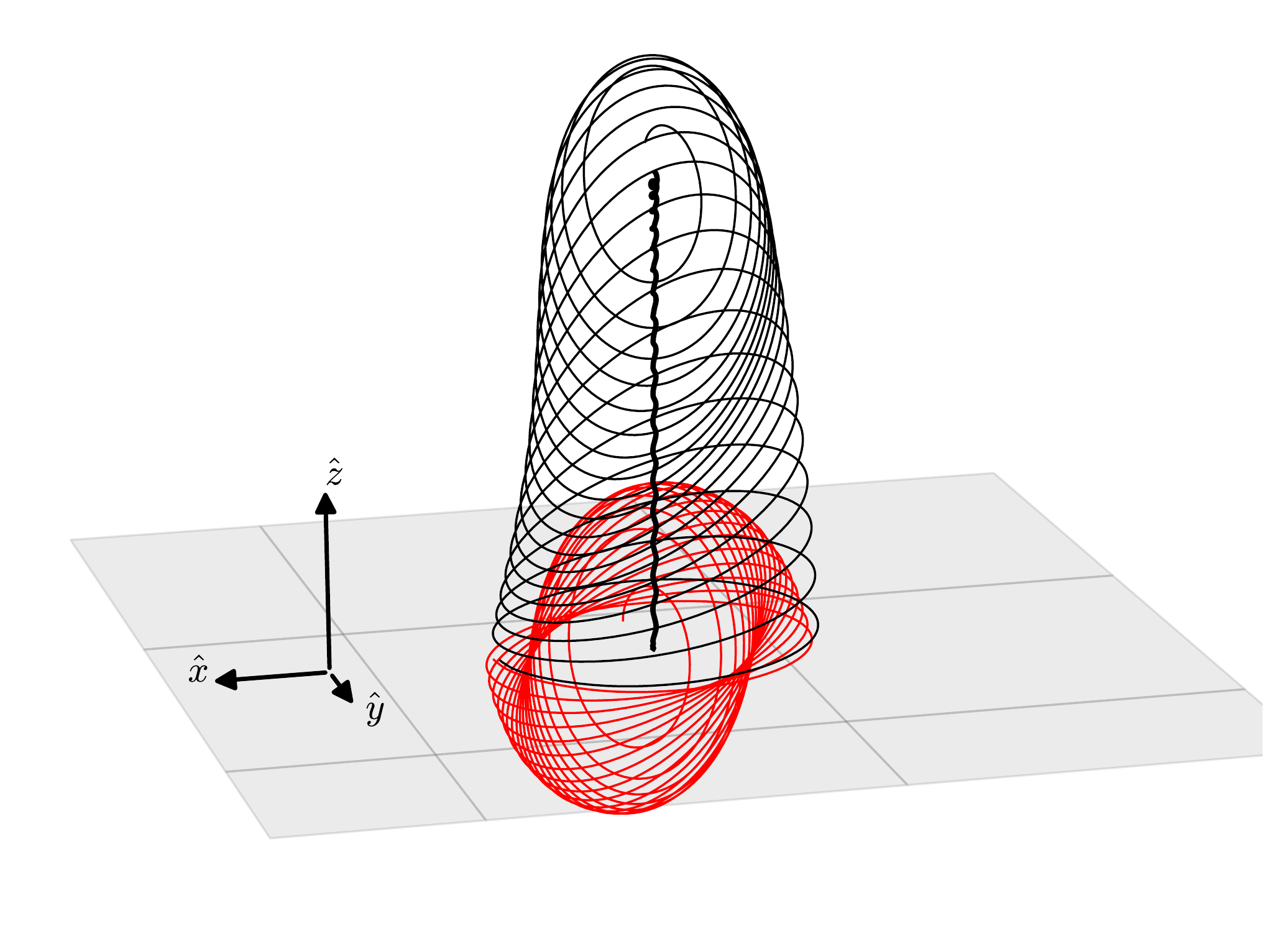}
    \includegraphics[width=0.49\linewidth]{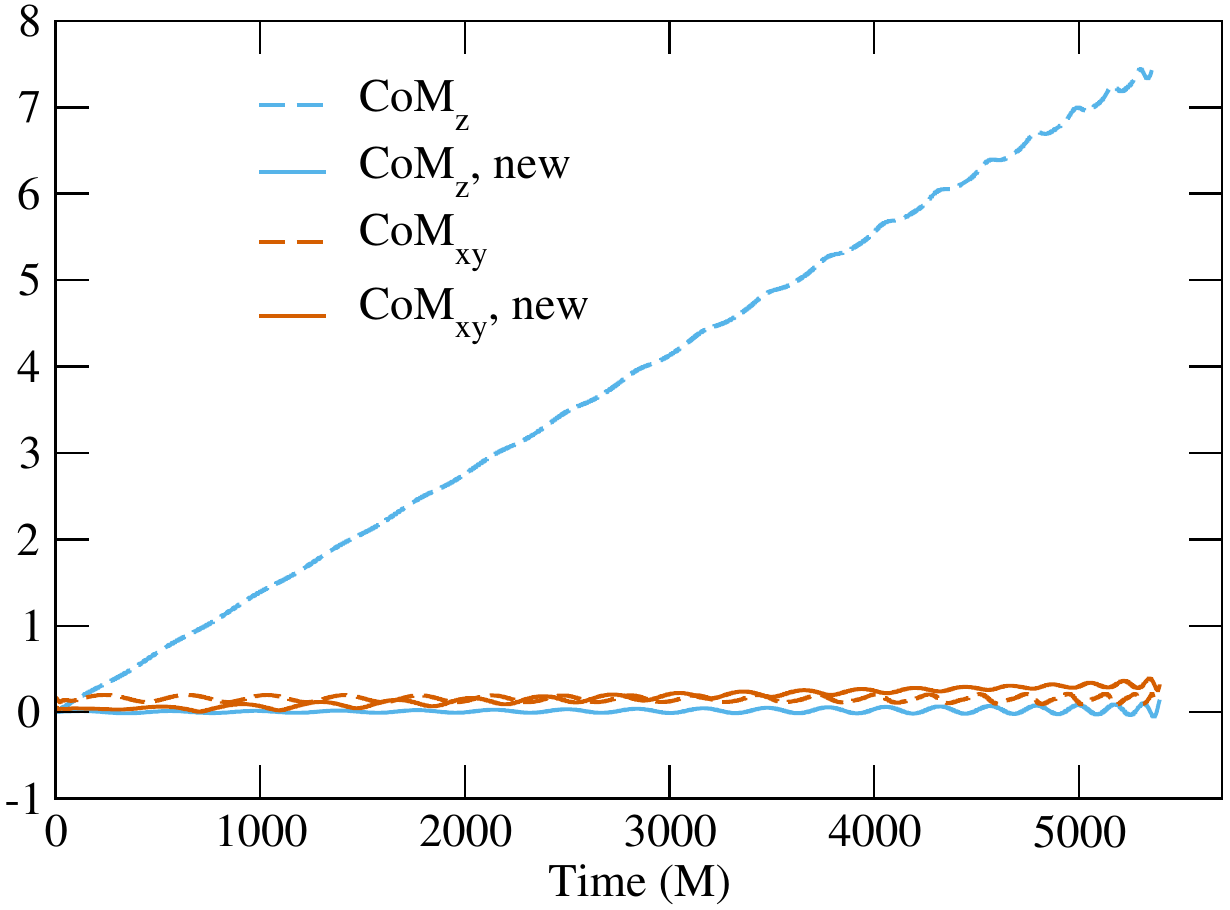}
  \end{center}
  \caption{ \label{fig:q3_Trajectories}{\bf Left}: The trajectories of
    the centres of the apparent horizons of the big black holes in the
    intertial frame. The thick black line represents the motion of the
    coordinate centre of mass for the drifting binary. The thin black
    curve correspond to initial data with large drift, the thick red,
    without. {\bf Right}: The components of the coordinate centre of
    mass for both runs. The dashed curves refer to initial data with
    large drift, the solid curves without.}
\end{figure}

Turning attention to the trajectories of the black holes, we find a
stark difference in the evolutions. Figure~\ref{fig:q3_Trajectories}
shows the motion of the large black hole in inertial coordinates for
both runs. The uncontrolled residual linear momentum $P_{\rm ADM}^z$
in the old initial data causes the centre of mass of the binary to
drift linearly during the evolution, as shown in the right panel of
Figure~\ref{fig:q3_Trajectories}.  Such a drift may have multiple
undesirable consequences.  Most immediately, it causes the
gravitational wave extraction spheres to be off-center from the
center-of-mass of the binary, which will cause mixing of the spherical
harmonic modes of the gravitational radiation, an effect discussed in
more detail below.  Moreover, {\tt SpEC}'s constraint preserving outer
boundary conditions~\cite{Rinne2008b,Rinne2008d,Buchman2007} are
designed to work best for low-order spherical harmonic modes of the
outgoing radiation.  If the binary is offset relative to the outer
boundary (for instance due to a drift of the center of mass), higher
order spherical harmonic components will become more important,
possibly leading to an additional runaway acceleration of the center
of mass~\cite{Szilagyi:2015rwa}.

\begin{figure}
   \begin{center}
     \parbox[b]{0.5\textwidth}{\includegraphics[width=0.47\textwidth]{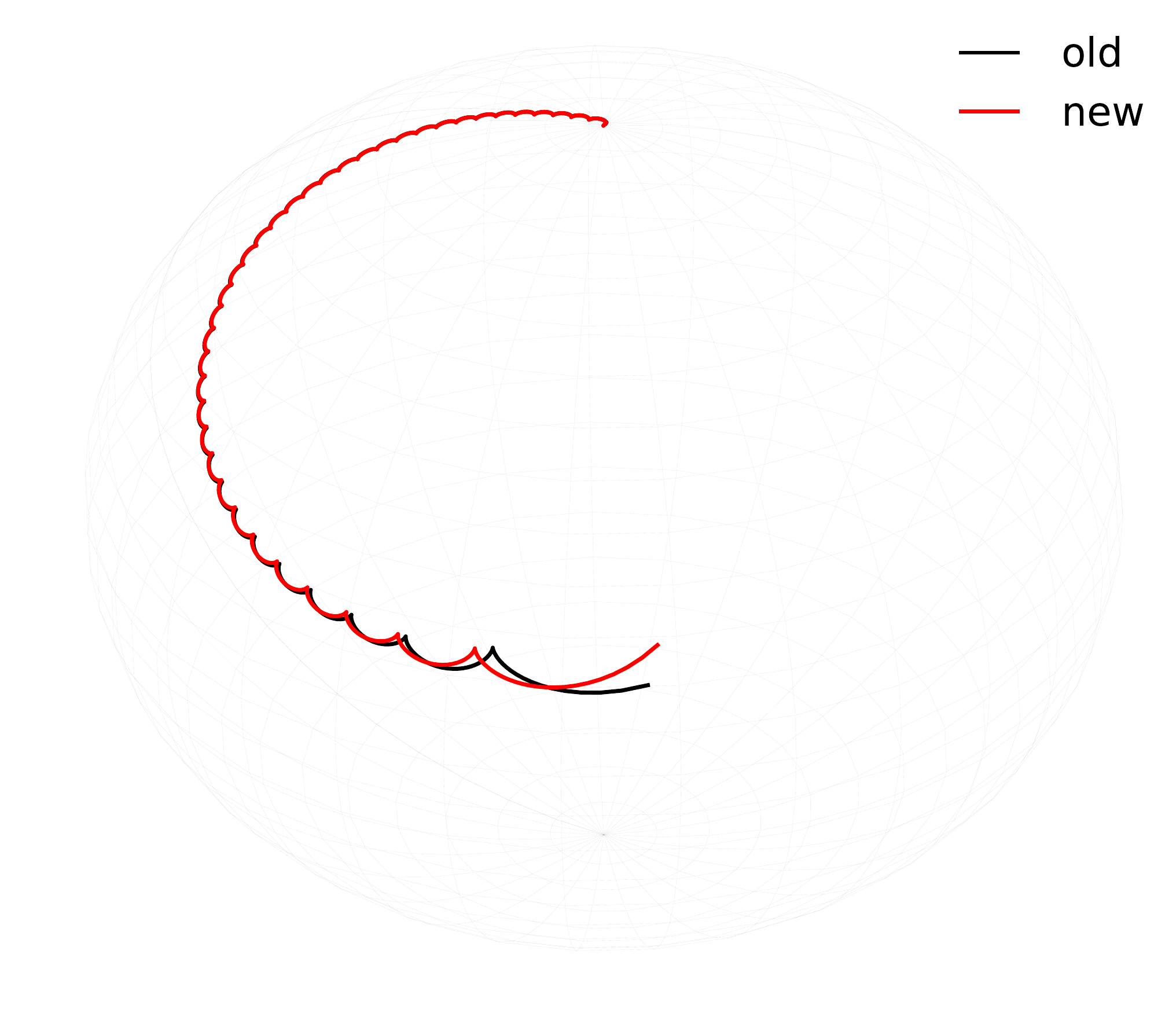}\\[.3cm]}
    \hfill
    \includegraphics[width=0.48\linewidth]{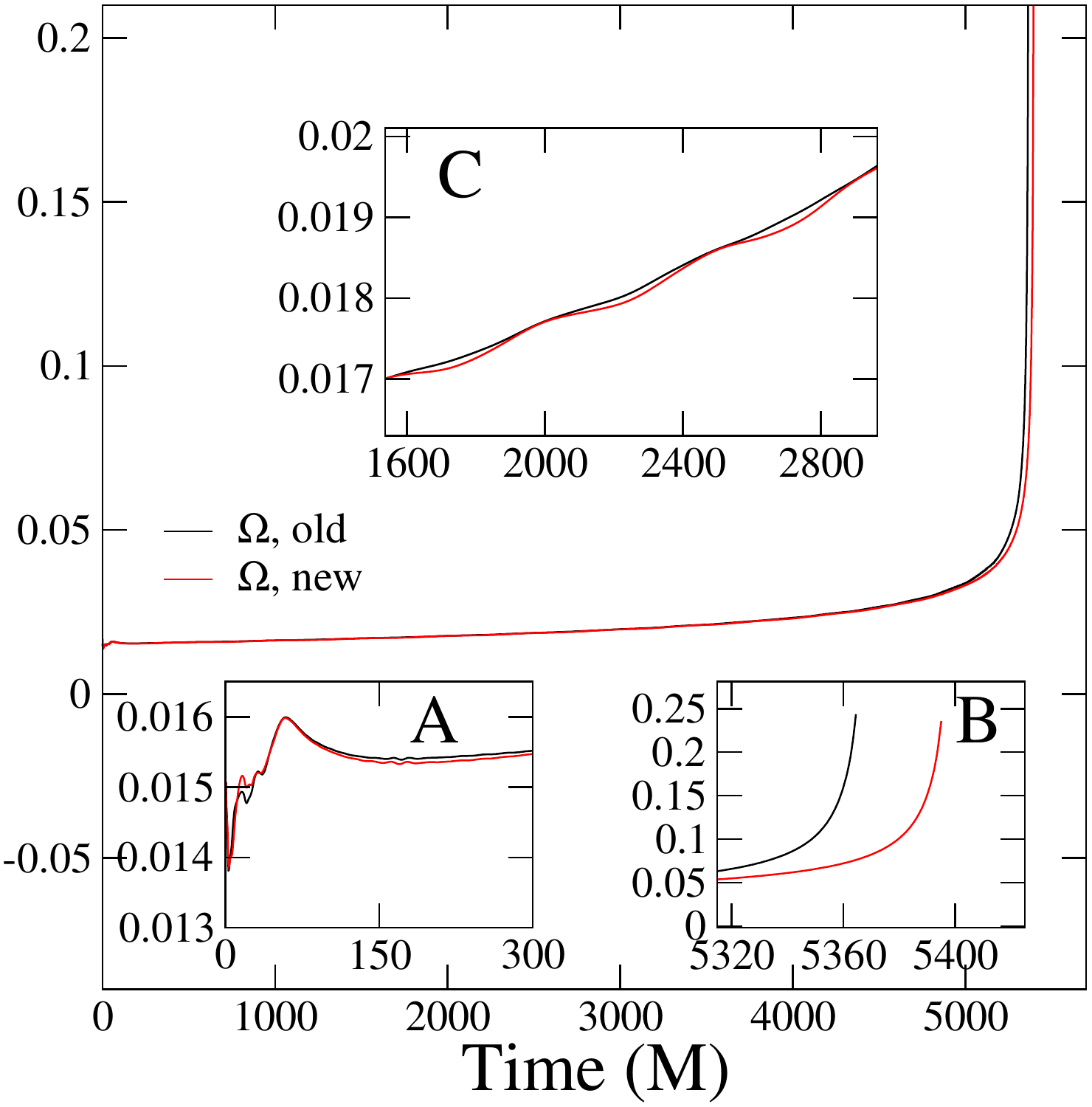}
  \end{center}
  \caption{\label{fig:q3_Omega}{\bf Left}: The direction
    $\hat{\w{\Omega}}(t)$ of the angular velocity vector on the unit
    sphere. Note the excellent agreement in precession dynamics. {\bf
      Right}: The magnitude $\Omega(t)$ of the angular velocity
    vector. The overall agreement is very good. Insets A and B
    highlight the different times to merger due to junk radiation
    dynamics, while inset C demonstrates the different
    eccentricities.}
\end{figure}

To examine the dynamics of the binary, we study the orbital frequency
vector \\${\w{\Omega}\equiv\w{r}\times\dot{\w{r}}/|\w{r}|^{2}}$. The left
panel of Figure~\ref{fig:q3_Omega} shows the projection of
$\w{\Omega}$ onto the unit sphere, making it apparent that the
precession and nutation dynamics are very similar until very close to
merger.  The right panel shows a plot of $\Omega\equiv|\w{\Omega}|$
from which several features are apparent. The evolution of $\Omega$ is
qualitatively the same in both cases, consistent with expectation that
removing a coordinate motion of the centre of mass does not change the
binary dynamics. Likewise the initial pulse of junk radiation (inset
A) appears quite similar.  However, small oscillations in $\Omega$ are
more pronounced in the new code (inset C). This is reflected in the
measured values of the eccentricities: $e=10^{-4}$ for the old,
$e=2.5\times10^{-4}$ for the new code.
The difference in eccentricity arises because the new term $v^i_o$
in the outer boundary condition Eq.~(\ref{eq:ShiftBC}) \emph{does}
slightly modify the content of the initial data.  In this particular
case, $|v_0|\sim
10^{-3}$, so that it is not unreasonable to expect the orbital
eccentricity to change by a comparable magnitude.  The initial orbital
frequency $\Omega_0$,
initial radial velocity $\dot
a_0$, and initial separation
$D_0$
listed in Table~\ref{tbl:Params} were tuned to result in essentially
vanishing eccentricity in the \emph{old} initial
data~\cite{Buonanno:2010yk}.  The new initial data constructed from
the identical initial data parameters must therefore have a slightly
larger eccentricity.  If we had tuned to vanishing eccentricity with
the \emph{new} initial data, then the old initial data would exhibit
the larger eccentricity.

The evolutions of the old and new initial data also result in a
different time to merger, cf. panel B of Fig.~\ref{fig:q3_Omega}.
This difference could be caused either by the slightly different
inspiral dynamics like eccentricity, or could simply be due to
truncation error of our low resolution evolution.

\begin{figure}
  \begin{center}
    \includegraphics{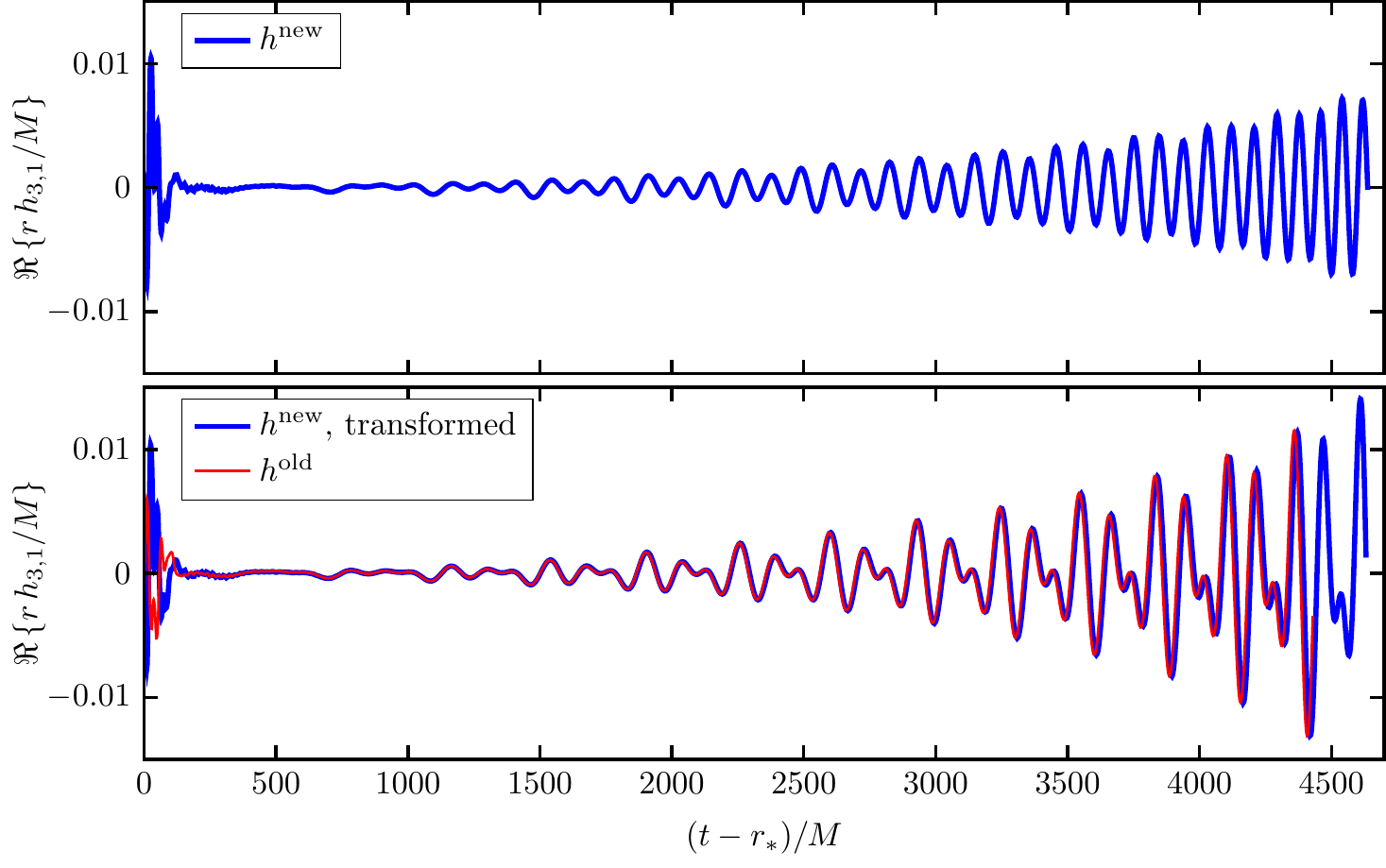}
  \end{center}
  \caption{ \label{fig:waveform_features} The $h_{3,1}$ waveform
      modes, as measured in evolutions of the original and the new
      initial data, and extrapolated to future null infinity,
      $\mathscr{I}^{+}$.  The upper panel shows the waveform $\hnew$
      from the new initial data, measured in a frame that is centered
      at the origin of the simulation coordinates.  The lower panel
      shows the same data with a transformation applied on
      $\mathscr{I}^{+}$ as described in the text, as well as the
      waveform $\hold$ from the original initial data measured in its
      simulation coordinates---in which the black holes are moving as
      shown in Fig.~\ref{fig:q3_Trajectories}.  Essentially, the
      center of mass is stationary at the origin in the upper panel,
      and is moving in the lower panel.
    }

\end{figure}

Finally, we examine the waveforms for the two runs. Most strikingly,
the movement of the coordinate centre of mass shown in
Figure~\ref{fig:q3_Trajectories} is also reflected in the
spherical-harmonic decomposition of the waveform.  This is most easily
seen in the sub-dominant modes. Figure~\ref{fig:waveform_features}
shows the $(\ell,m) = (3,1)$ modes of the spin-weighted
spherical-harmonic (SWSH) decompositions of the waveforms $\hold$
measured from the old initial data and $\hnew$ measured from the new
initial data.  Since gravitational waves in {\tt SpEC} are extracted
on a coordinate sphere centered on the origin, a drifting source mixes
the modes of the SWSH decomposition.  As seen in the lower panel of
the figure, this mixing introduces very large effects.  To verify that
these effects are primarily due to the motion of the center of mass,
we have also transformed $\hnew$ to a frame in which the center of
mass is moving as in the original initial data.  The initial position
of $\hnew$ is transformed to agree with the center of mass of the old
initial data as measured by Eq.~(\ref{eq:CoM}), and its velocity is
transformed to agree with $\w{P}_{\text{ADM}} / M_{\text{ADM}}$ of the
old initial data as measured by Eq.~(\ref{eq:PADM}).  This
transformation is applied entirely at future null infinity by the
method described in~\cite{Boyle2015a}, and is a special case of a BMS
transformation~\cite{Sachs1962,Newman1962}.  It will thus be seen in
any waveforms, whether extrapolated~\cite{Boyle-Mroue:2008} (as seen
here) or extracted by Cauchy characteristic
methods~\cite{Bishop1996,Babiuc2005,Taylor:2013zia,Handmer:2015}.  As
shown in the lower panel of Fig.~\ref{fig:waveform_features}, the
transformation reproduces the features seen in $\hold$ very well.

Mode decompositions like this one are used very frequently for
analyzing numerical models, and for constructing analytical models.
If they are unmodeled and uncontrolled, effects like those seen in the
lower panel will simply appear to be errors in the waveform.  This
could negatively impact uncertainty estimates of numerical
simulations, error estimates for analytical waveforms, or calibration
of waveform models to numerical results.  These effects will also be
present in any calculation that uses the waveforms to compute physical
quantities such as the flux of linear and angular momentum. %
By removing extraneous displacements and boosts, this new initial data
code simplifies such analyses.\footnote{The drift described here is a
  linear motion due to residual linear momentum in initial
  data. Controlling this drift will not help for other types of motion
  present in very long simulations; see~\cite{Szilagyi:2015rwa}.}

%%%%%%%%%%%%%%%%%%%%%%%%%%%%%%%%%%%%%%%%%%%%%%%%%%%%%%%%%%%%%%%%
%%%%%%%%%%%%%%%%%%%%%%%%%%%%%%%%%%%%%%%%%%%%%%%%%%%%%%%%%%%%%%%%
\section{Discussion}
\label{sec:Discussion}
%%%%%%%%%%%%%%%%%%%%%%%%%%%%%%%%%%%%%%%%%%%%%%%%%%%%%%%%%%%%%%%%
%%%%%%%%%%%%%%%%%%%%%%%%%%%%%%%%%%%%%%%%%%%%%%%%%%%%%%%%%%%%%%%%
Numerical evolution of binary black hole spacetimes requires accurate
initial data. In this work we have improved the initial data
techniques in {\tt SpEC} to allow access to a much wider parameter
space of generically precessing high mass ratio, high-spin binaries. A
more flexible domain decomposition allows for stable solution for
high-mass ratio and high spin binaries. An enhanced root-finding
algorithm is used to achieve desired physical parameters for the
binary. This becomes important when a naive analytic Jacobian is not
appropriate, which is precisely the case for high mass ratios and
spins, see Figure~\ref{fig:RootFinding}. Adaptive mesh refinement
drastically improves efficiency and robustness of the code, displaying
exponential convergence of constraints,
c.f. Figure~\ref{fig:Many_Convergence}. Finally, a new method to
control the linear momentum is used to eliminate a linear drift of the
centre of mass during evolution. This in turn nullifies spurious
gravitational mode mixing, which is of paramount importance for
construction of hybrid waveforms or calibration of phenomenological
models as demonstrated by Figure~\ref{fig:waveform_features}.

\begin{figure}
  \begin{center}
    \includegraphics[width=0.48\linewidth,trim=80 0 50 30,clip=True]{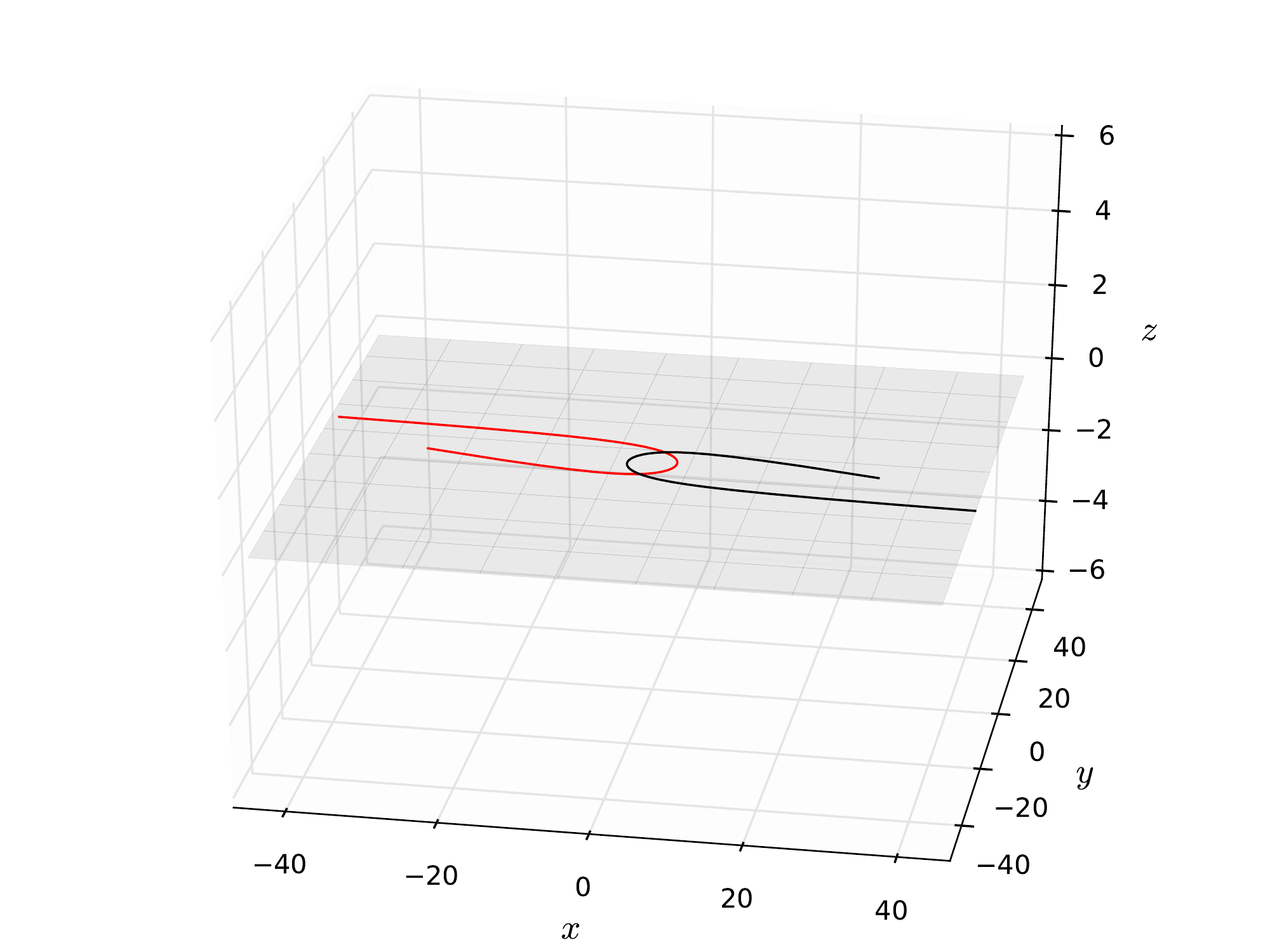}
\hfill
    \includegraphics[width=0.48\linewidth,trim=80 0 50 30,clip=True]{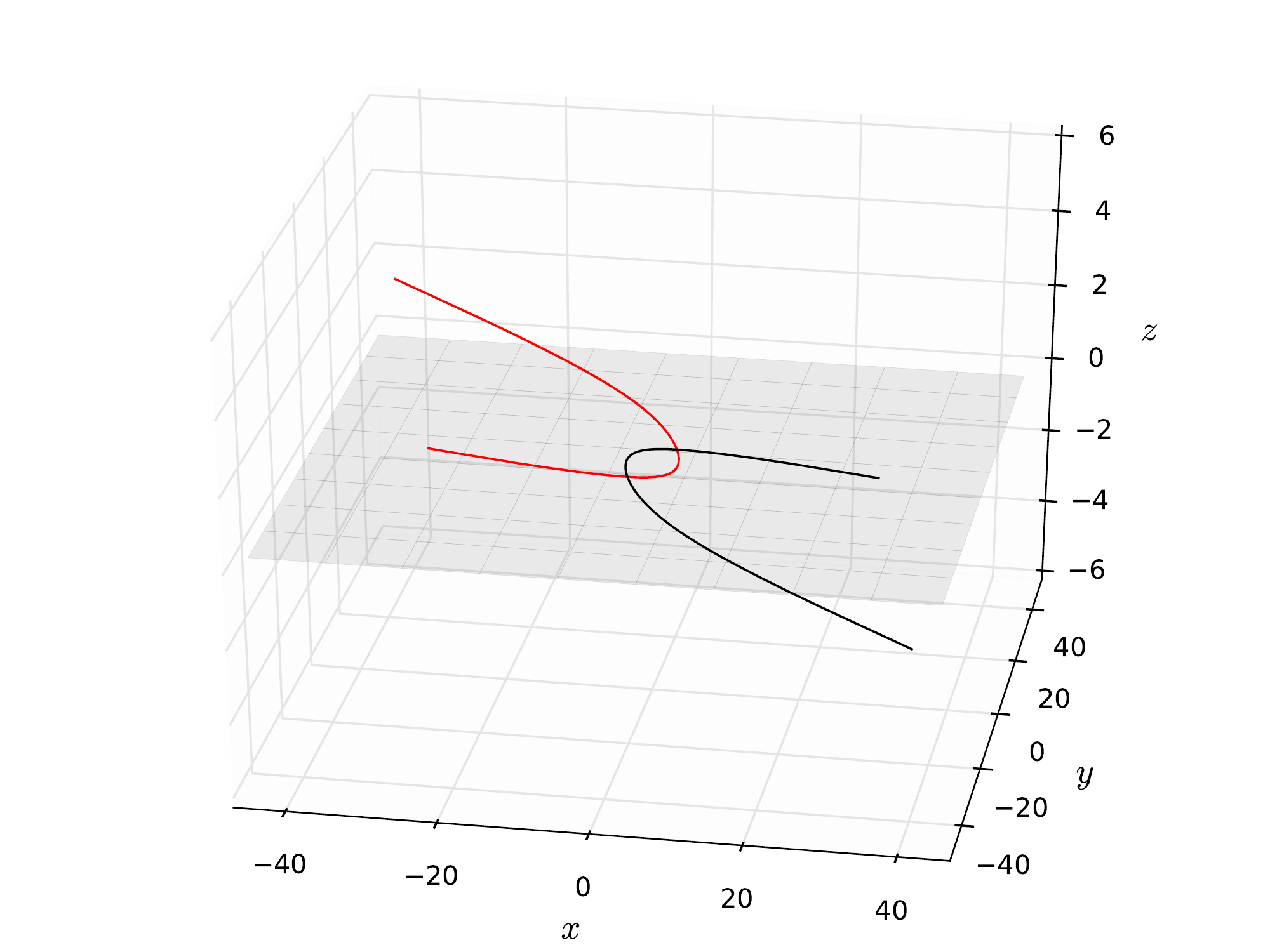}
  \end{center}
  \caption{ \label{fig:Traj_q1} Hyperbolic encounter of two equal mass
    black holes shown through the coordinate trajectories of the black
    holes.  {\bf Left}: non-spinning black holes. {\bf Right}: black
    holes with spin $\w{\chi}_{1}=\w{\chi_{2}}=(0.5,0.0,0.0)$. The
    black holes start on the $x$-axis in the $x-y$ plane (shown in
    grey). In the spinning case the motion is not confined to this
    plane.}
\end{figure}

An interesting application of the improved initial data code is the
construction of initial data for hyperbolic encounters. Such systems
have been studied in the past
(e.g.~\cite{Damour:2014afa,Sperhake2009,Healy2008}) and provide a
laboratory for exploring strong field physics in a different regime
than the binary inspiral. Using the new code, we have successfully
constructed initial data for hyperbolic encounters for a selection of
mass ratios and spins, which was not possible before in {\tt SpEC}. As a
simple example, we evolve two systems of two equal mass black holes
that are initially separated by 60M and have a velocity of
$\approx0.14c$.  Both systems have the same impact parameter
$b_{NR}=15M$, and differ only in the black hole spins: In one case the
black holes are non-spinning, in the other both holes have
dimensionless spins $\chi=0.5$ initially in the $x$ direction.
Figure~\ref{fig:Traj_q1} shows the trajectories of the two black
holes.  In the presence of spin, the spin-orbit interactions cause the
plane of scattering to change and also change the deflection angle of
the hyperbolic encounter. Exploration of other parameters is left to
future investigations.

Another application is the construction of initial data for binaries
with very small initial separation, corresponding to only a few orbits
before merger. This is useful if one is interested in the properties
of the merger remnant, e.g. for calibrating analytical waveform models
but evolving a long inspiral is too computationally expensive. As an
example, we construct initial data for a system with $q=21$,
$\chi_{1}=0.66$, $\chi_{2}=0.41$ (oriented in random directions) and
initial orbital frequency of $M\Omega=0.032$, and initial coordinate
separation $D_{0}=8.82$ M. We note that initial data for binaries
near ISCO at high mass-ratio is challenging and further work remains
to be done to make it robust for $q>10$ regime.

%%%%%%%%%%%%%%%%%%%%%%%%%%%%%%%%%%%%%%%%%%%%%%%%%%%%%%%%%%%%%%%%
\section*{Acknowledgements}
%%%%%%%%%%%%%%%%%%%%%%%%%%%%%%%%%%%%%%%%%%%%%%%%%%%%%%%%%%%%%%%%

We thank Geoffrey Lovelace, Larry Kidder and Mark Scheel for helpful
discussions.  Calculations were performed with the {\tt
  SpEC}-code~\cite{SpECwebsite}.  We gratefully acknowledge support
from NSERC of Canada, from the Canada Research Chairs Program, and
from the Canadian Institute for Advanced Research.  FF gratefully
acknowledges support from the Vincent and Beatrice Tremaine
Postdoctoral fellowship at CITA. Support for this work was provided by
NASA through Einstein Postdoctoral Fellowship grant numbered
PF4-150122.  We further gratefully acknowledge support from the
Sherman Fairchild Foundation; from NSF Grants PHY-1306125 and
AST-1333129 at Cornell; and from NSF Grants No.~PHY-1440083 and
AST-1333520 at Caltech.  Calculations were performed at the Gravity
cluster and the GPC supercomputer at the SciNet HPC
Consortium~\cite{scinet}; SciNet is funded by: the Canada Foundation
for Innovation (CFI) under the auspices of Compute Canada; the
Government of Ontario; Ontario Research Fund (ORF) -- Research
Excellence; and the University of Toronto.  Further calculations were
performed on the Briar\'{e}e cluster from Sherbrooke University,
managed by Calcul Qu{\'e}bec and Compute Canada. The operation of this
supercomputer is funded by the Canada Foundation for Innovation (CFI),
Minist\`{e}re de l'\'{E}conomie, de l'Innovation et des Exportations
du Qu\'{e}bec (MEIE), RMGA and the Fonds de recherche du Qu\'{e}bec -
Nature et Technologies (FRQ-NT).\\

%%%%%%%%%%%%%%%%%%%%%%%%%%%%%%%%%%%%%%%%%%%%%%%%%%%%%%%%%%%%%%%%%%%%%%%%%%%%%%%
\section*{References}
%%%%%%%%%%%%%%%%%%%%%%%%%%%%%%%%%%%%%%%%%%%%%%%%%%%%%%%%%%%%%%%%%%%%%%%%%%%%%%%
\bibliographystyle{unsrt}
\bibliography{References/References}
%\begin{thebibliography}{99}
%\end{thebibliography}

\end{document}